\documentclass[twocolumn]{aastex631}

\usepackage{CJK}

\newcommand{\beq}{\begin{equation}}
\newcommand{\eeq}{\end{equation}}

\shorttitle{WR Star Spin}
\shortauthors{Ma \& Fuller}

\graphicspath{{./}{figures/}}

\begin{document}
\begin{CJK*}{UTF8}{gbsn}

\title{Tidal Spin-up of Black Hole Progenitor Stars}
\correspondingauthor{Linhao Ma}
\email{lma3@caltech.edu}

\author[0000-0001-6117-5750]{Linhao Ma（马林昊）}
\affiliation{TAPIR, Mailcode 350-17, California Institute of Technology, Pasadena, CA 91125, USA}

\author[0000-0002-4544-0750]{Jim Fuller}
\affiliation{TAPIR, Mailcode 350-17, California Institute of Technology, Pasadena, CA 91125, USA}

\begin{abstract}

Gravitational wave observations indicate the existence of merging black holes (BHs) with high spin ($a\gtrsim0.3$), whose formation pathways are still an open question. A possible way to form those binaries is through the tidal spin-up of a Wolf--Rayet (WR) star by its BH companion. In this work, we investigate this scenario by directly calculating the tidal excitation of oscillation modes in WR star models, determining the tidal spin-up rate, and integrating the coupled spin--orbit evolution for WR--BH binaries. We find that for short-period orbits and massive WR stars, the tidal interaction is mostly contributed by standing gravity modes, in contrast to Zahn's model of travelling waves which is frequently assumed in the literature. The standing modes are less efficiently damped than traveling waves, meaning that prior estimates of tidal spin-up may be overestimated. We show that tidal synchronization is rarely reached in WR--BH binaries, and the resulting BH spins have $a \lesssim 0.4$ for all but the shortest period ($P_{\rm orb} \! \lesssim 0.5 \, {\rm d}$) binaries. Tidal spin-up in lower-mass systems is more efficient, providing an anti-correlation between the mass and spin of the BHs, which could be tested in future gravitational wave data. Nonlinear damping processes are poorly understood but may allow for more efficient tidal spin-up. We also discuss a new class of gravito-thermal modes that appear in our calculations.

\end{abstract}


\section{Introduction}
\label{sec:introduction}

The spins of stellar-mass black holes (BHs) are still not fully understood. Most BHs detected from LIGO/Virgo events have low aligned components of their spins \citep{Abbott2019,Zaldarriaga2018,Miller2020,Roulet2021,Zevin2021}, which agrees with predictions of efficient angular momentum (AM) transport within the interiors of massive stars. Such processes remove the majority of AM from the stellar core, predicting slowly rotating remnants after core-collapse \citep{fuller:19,ma:19}. These theories are approximately consistent with core-rotation rate measurements of low-mass red giants from asteroseismology \citep{beck:12,mosser:12,deheuvels:14,Deheuvels2015,Triana2017,Gehan2018}, with a few discrepancies \citep{Eggenberger:19}. Yet, among a small fraction of LIGO/Virgo BHs and a majority of high-mass X-ray binaries \citep{Miller2015}, high BH spins are measured. Therefore it still remains a theoretical challenge to explain the existence of these rapidly rotating objects (see, e.g. discussions in \citealt{Qin2019,Belczynski2021,Fishbach2022}).

A natural scenario to form high-spin BHs is through binary interactions, as nearly all BHs with spin measurements are found via BH mergers or X-ray binaries. One possible progenitor of BH binaries are Wolf--Rayet--BH binaries. Such a system is formed from an ordinary massive binary system, where the primary collapses to a (likely slowly rotating) BH, and then strips off the envelope of the secondary, making it a Wolf--Rayet (WR) star. Tidal interactions during the WR phase could possibly spin up the latter, forming a rapidly spinning BH. Many studies have investigated this scenario and made predictions for the spins of the second-born BHs \citep{Kushnir2017,Qin2018,Bavera2020,Belczynski2020,Olejak2021,fuller2022}, finding they can be large for sufficiently close binary systems ($P_{\rm orb} \lesssim 1 \, {\rm day}$).

However, in most of these studies, the tidal response of the WR star to the BH companion is not calculated directly. Instead, an effective tidal torque calculated from Zahn's theory of dynamical tides (\citealt{zahn:75,zahn:77}, see also \citealt{goldreich:89}) is often assumed. The basic picture of Zahn's theory is as follows: gravity waves are tidally excited near the convective core-radiative envelope interface inside a star. The waves propagate outwards and damp due to radiative diffusion near the surface of the star. The damping is often so strong that the waves dissipate before reaching the surface and behave as travelling waves rather than standing waves. The energy and AM deposited by the waves can be calculated and translated to an effective tidal torque. While this picture is often assumed in studies of tidally excited waves, it has not been closely examined in binaries involving a WR star.

In this work, we directly solve for oscillation modes of WR stellar models, quantifying their tidal coupling strengths and dissipation rates. We then compute AM transfer rates and model their spin evolution and resulting BH spins, comparing to those from Zahn's theory. The plan of this paper is as follows: in \S \ref{sec:formalism} we review the basic formalism of dynamical tides for calculating tidal torques based on stellar evolution models, and we summarize the setups of our models of the WR stars; in \S \ref{sec:results-modes} and \S \ref{sec:results-wr} we present our analysis for the tidally excited modes and the stellar spin evolution. We discuss our results in \S \ref{sec:discussion} and conclude in \S \ref{sec:conclusion}.

\section{Tidal Torques by Dynamical Tides}
\label{sec:formalism}

\begin{table}
    \centering
    \begin{tabular}{ccccc}
         \\\hline
         model & $M_\mathrm{ZAMS}$ & $M_\mathrm{WR}$ & Dutch factor & desired $Z$ \\
         \hline
         1 & $15\,M_\odot$ & $3\,M_\odot$ & 0.5 & $10^{-2}Z_\odot$ \\
         2 & $20\,M_\odot$ & $5\,M_\odot$ & 0.5 & $10^{-2}Z_\odot$ \\
         3 & $30\,M_\odot$ & $10\,M_\odot$ & 0.5 & $10^{-2}Z_\odot$ \\
         4 & $45\,M_\odot$ & $18\,M_\odot$ & 0.5 & $10^{-2}Z_\odot$ \\
         5 & $60\,M_\odot$ & $27\,M_\odot$ & 0.5 & $10^{-2}Z_\odot$ \\
         6 & $80\,M_\odot$ & $38\,M_\odot$ & 0.5 & $10^{-2}Z_\odot$ \\
         7 & $100\,M_\odot$ & $50\,M_\odot$ & 0.5 & $10^{-2}Z_\odot$ \\
         8 & $120\,M_\odot$ & $62\,M_\odot$ & 0.5 & $10^{-2}Z_\odot$ \\
         9 & $15\,M_\odot$ & $3\,M_\odot$ & 4.0 & $Z_\odot$ \\
         10 & $20\,M_\odot$ & $5\,M_\odot$ & 4.0 & $Z_\odot$ \\
         11 & $30\,M_\odot$ & $10\,M_\odot$ & 4.0 & $Z_\odot$ \\
         12 & $45\,M_\odot$ & $18\,M_\odot$ & 4.0 & $Z_\odot$ \\
         13 & $60\,M_\odot$ & $26\,M_\odot$ & 3.0 & $Z_\odot$ \\
         14 & $80\,M_\odot$ & $38\,M_\odot$ & 2.0 & $Z_\odot$ \\
         15 & $100\,M_\odot$ & $49\,M_\odot$ & 1.7 & $Z_\odot$ \\
         16 & $120\,M_\odot$ & $61\,M_\odot$ & 1.5 & $Z_\odot$ \\
         \hline
    \end{tabular}
    \caption{Parameters of our Wolf--Rayet star models. We fixed the metallicities of all models to $Z=0.01\,Z\odot$ and adapted their Dutch wind scaling factors to match the mass-loss rates for the desired metallicities. See discussions in the main text.}
    \label{tab:models}
\end{table}

In classical tidal theory, tides can be decomposed into two components: equilibrium tides and dynamical tides. The former corresponds to the global distortion of the star, while the latter is composed of internal oscillations, which is believed to be a dominant cause of tidal dissipation. From \cite{ma2021}, the energy dissipation rate of a tidally forced oscillation mode $\alpha$ excited by the tidal potential of an aligned and circular orbiting secondary is given by
\beq
\label{eq:tide_dissip}
\dot{E}_\alpha = \frac{m\omega_\alpha \Omega_\mathrm{orb}\gamma_\alpha q^2M_1R_1^2|W_{lm}Q_\alpha|^2\omega_\mathrm{f}^2}{(\omega_\alpha-\omega_\mathrm{f})^2+\gamma_\alpha^2}\bigg(\frac{R_1}{a}\bigg)^{2(l+1)}\,,
\eeq
where $\omega_\alpha$ and $\gamma_\alpha$ are the mode frequency and damping rate, and $\omega_\mathrm{f} = m(\Omega_{\rm orb}-\Omega_{\rm spin})$ is the tidal forcing frequency (measured in the frame co-rotating with the primary), and $\Omega_{\rm spin}$ is the star's angular rotation frequency. $M_1$ and $R_1$ are the mass and radius of the primary, $q=M_2/M_1$ is the mass ratio of the secondary to the primary, $a$ and $\Omega_\mathrm{orb}$ are the semi-major axis and the angular frequency of the orbit. $l$ and $m$ are the mode's angular and azimuthal wave numbers and $W_{lm}$ is an expansion coefficient of the tidal potential. $Q_\alpha\equiv\langle\xi_\alpha|\nabla(r^lY_{lm})\rangle/\omega_\alpha^2$ is the dimensionless overlap integral describing the spatial coupling between the mode and the tidal potential, which is calculated by the relation $Q_\alpha = -(2l+1)\delta\Phi_\alpha/(4\pi\omega_\alpha^2)$ \citep{fullerheartbeat:17}, where $\delta\Phi_\alpha$ is the surface gravity potential perturbation. The mode angular momentum dissipation rate is related to the energy dissipation by \citep{fullerheartbeat:17}
\beq
\label{eq:j_dissip}
\dot{J}_\alpha = \frac{\dot{E}_\alpha}{\Omega_\mathrm{orb}}\, ,
\eeq
assuming a circular orbit. Hence, by solving for the internal oscillation modes (with $\omega_\alpha$, $\gamma_\alpha$ and $\mathcal{Q}_\alpha$) inside the primary, we are able to calculate the energy dissipation and tidal spin-up rate, given a companion mass and orbit.

\subsection{Stellar Models}
\label{sec:models}

We built our WR star models with the MESA stellar evolution code \citep{paxton:11,paxton:13,paxton:15,paxton:18,paxton:19}. Instead of using the binary options in MESA, we construct our models as follows: we start with a number of zero-age main-sequence (ZAMS) single star models with a variety of masses, summarized in Table \ref{tab:models}. The stars evolve to core hydrogen depletion before the stripping-off process occurs. We simulate this process by artificially removing the outer hydrogen envelope immediately after hydrogen depletion (defined by the time when the central hydrogen fraction drops below $10^{-5}$), producing a helium star as the initial setup for the WR star. We then restart the evolution until the end of core helium depletion (when the central helium fraction drops below $10^{-5}$), and we output the stellar pulsation parameters to be used later for spin-evolution calculations. Example MESA inlists are available on Zenodo under an open-source 
Creative Commons Attribution license: 
\dataset[doi:10.5281/zenodo.7935443]{https://doi.org/10.5281/zenodo.7935443}, and the model parameters are summarized in Table \ref{tab:models}.

During the helium burning phase, we compute the internal oscillations of the models with the GYRE stellar oscillation code \citep{townsend:13,townsend2018,goldstein2020}. We use the second order Magnus differential scheme to calculate non-adiabatic modes, as it proves to be the most reliable when dealing with low-frequency oscillations. We specify our search to $l=m=2$ modes since this is the dominant part of the tidal potential in aligned and circular orbits, with the corresponding $W_{22}=\sqrt{3\pi/10}$. Example GYRE inputs are available on Zenodo under an open-source 
Creative Commons Attribution license: 
\dataset[doi:10.5281/zenodo.7935443]{https://doi.org/10.5281/zenodo.7935443}. Once we have the mode solutions, we integrate the spin--orbit evolution with Eq. \ref{eq:tide_dissip} and \ref{eq:j_dissip}, summing over all modes. We assume the primary remains rigidly rotating during the evolution, due to the strong AM diffusion inside WR stars \citep{fuller2022}. We use our non-rotating mode solutions all along in the integration, as we will see that most systems never get to tidal synchronization, such that the rotational effects can be ignored.

An important process related to the spin--orbit evolution is the large wind mass loss experienced by WR stars (e.g., \citealt{sander:20}), which removes AM from both the spin and the orbit. The mass loss rates of high-mass WR stars are somewhat uncertain \citep{sander:20,Vink2022}, especially at low-metallicity, hence there are few reliable observed/modelled values to compare with. We simulate the mass loss with the ``Dutch'' wind scheme \citep{nugis:00} in MESA with $\eta=0.5$ and include its effects in our integration. The mass-loss rate has a strong dependence on the metallicity of the star. However, we find that GYRE was unable to solve the oscillations correctly for some of our massive models at solar metallicity due to MESA's artificial treatment of super-Eddington near-surface layers. We hence used a universal metallicity $Z=0.01\,Z_\odot$ in all our models so that the stellar structure can be more accurately modeled. Oscillations solved from these models are reasonable approximations since the mode properties are mostly determined by the deep internal structure of the stars which are not strongly dependent on metallicity.

To estimate the evolution and mass loss rates of higher metallicity stars, we increase the wind scaling factor to match the mass-loss rate of an alternative model with the desired metallicity. For instance, to simulate a $10 \, M_\odot$ WR star at solar metallicity, we use a wind scaling factor of $\eta=4$ for our $Z=0.01 \, Z_\odot$ model of the same mass, which produces roughly the same mass loss rate as a $Z=Z_\odot$ model with $\eta=0.5$. In the following, we will simply reference the models with their desired metallicity, yet the readers should keep in mind that the underlying models actually have $Z=0.01\,Z_\odot$ and adapted Dutch factors, which are summarized in Table \ref{tab:models}.

\begin{figure*}
    \centering
    \includegraphics[width=\textwidth]{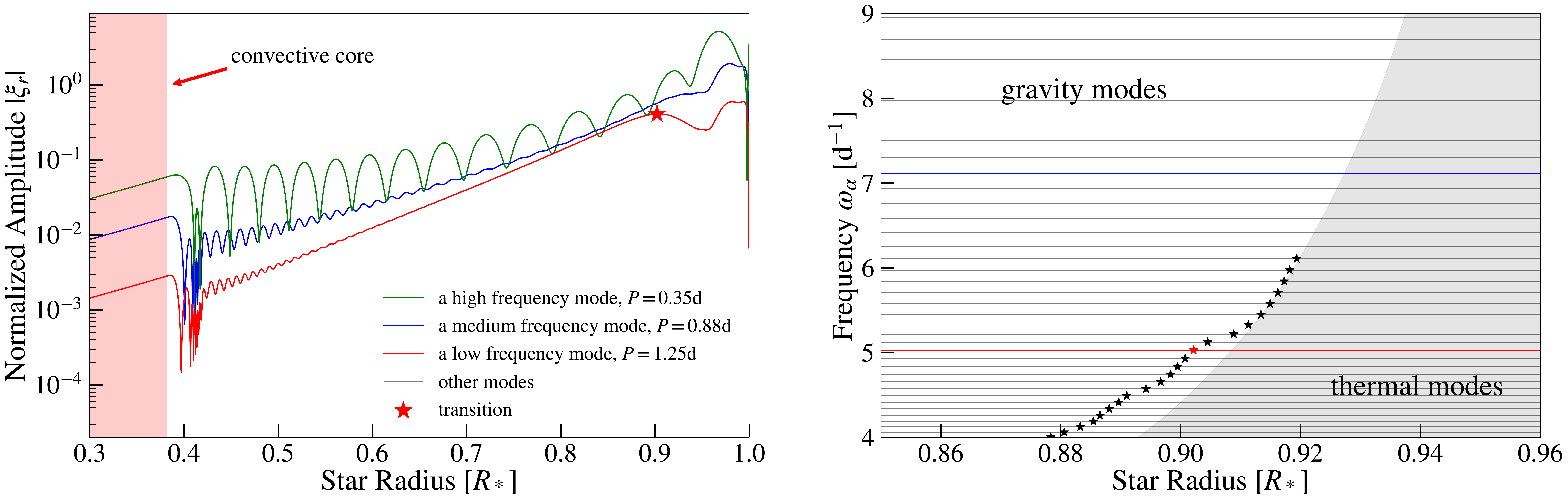}
    \caption{{\bf Left}: Example mode eigenfunctions for a $10\,M_\odot$ Wolf--Rayet star model at solar metallicity during helium burning. For high-frequency modes, we see a standing g-mode (green line) excited near the convective core (red region) boundary, in contrast with Zahn's assumption of travelling waves. As the frequency decreases, the modes become travelling gravity waves (blue line), damping near the surface (Zahn's formalism). When the frequency continues to decrease, the modes become mixed modes with a travelling g-mode component and a thermal mode component. The red star marks the transition point, calculated by the local maximum of the eigenfunction. {\bf Right}: A detailed look at the transition points between g-modes and thermal modes (stars). The lines show the frequencies of all modes and the colored lines correspond to the example modes in the left panel. The transition points agree well with the theoretically derived ones where $\omega_\alpha=\omega_\mathrm{crit}$ (the boundary between two propagation regions, cf. Appendix \ref{app:dispertion}). At higher frequency the thermal mode region (shaded) gets narrower and disappears.}
    \label{fig:eigenfunction}
\end{figure*}

\begin{figure}
    \centering
    \includegraphics[width=\columnwidth]{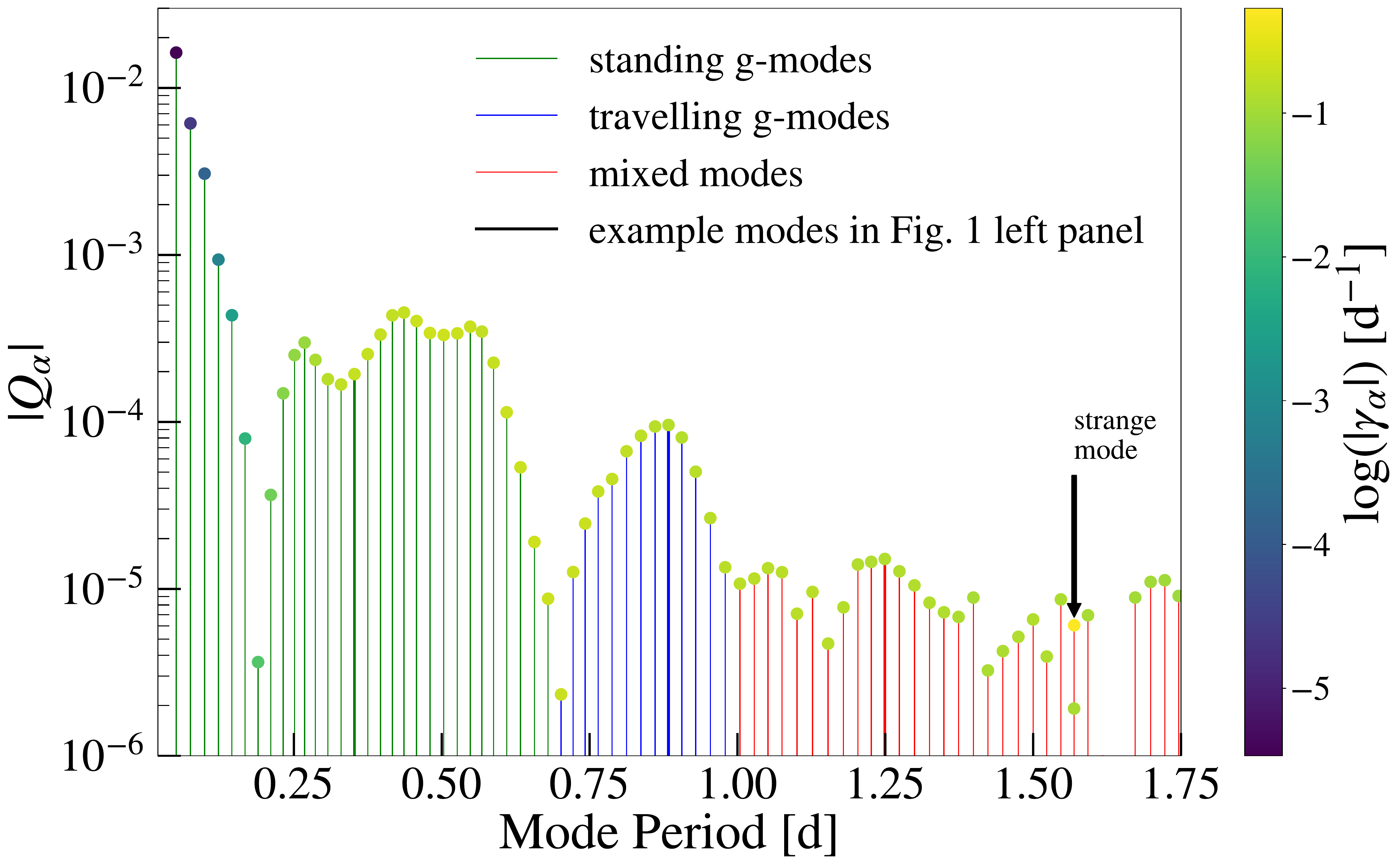}
    \caption{The overlap integral $|Q_\alpha|$ as a function of mode period from a  $10\,M_\odot$ Wolf--Rayet star model at solar metallicity during helium burning (same as Figure \ref{fig:eigenfunction}). Circle colors indicate mode damping rates $|\gamma_\alpha|$, while line color indicates the mode type (the thick lines show the corresponding modes in Figure \ref{fig:eigenfunction} left panel). ``Strange mode'' solutions often appear at long periods with excess damping rates and unusual period spacings (see Appendix \ref{app:strange_modes}).}
    \label{fig:modes}
\end{figure}

\section{Mode Morphology}
\label{sec:results-modes}

Figure \ref{fig:eigenfunction} shows some example mode eigenfunctions from a $10\,M_\odot$ WR star model at solar metallicity during the helium burning phase. We see that there is a distinction between the high-frequency ($P\lesssim0.8\,\mathrm{d}$, green line) and low-frequency ($P\gtrsim0.8\,\mathrm{d}$, blue and red lines) modes. For a typical high-frequency mode, the eigenfunction appears to be a standing low-order gravity wave trapped in the radiative envelope of the star. This is in contrast with the \cite{zahn:75} model for travelling waves that damp near the stellar surface. When the tidal forcing frequency $\omega_\mathrm{f}$ becomes close to the frequency of one of these modes, a resonance occurs and the energy/angular momentum dissipation becomes dominated by it (cf. Equation \ref{eq:tide_dissip}). The tidal torque contributed by this standing mode is different from Zahn's theory, and we will see in \S \ref{sec:results-wr} that Zahn's results on spin--orbit evolution are significantly altered.

At lower frequency, the modes turn to travelling waves (blue line) as Zahn assumed, since the g-mode dispersion relation indicates an imaginary wave number ${\rm Im}(k_r)\propto1/\omega_\alpha^2$ (see Appendix \ref{app:dispertion}), i.e., the spatial evanescence becomes larger at lower frequency. For even lower frequency, the eigenfunction becomes a mixed mode which can be separated into two components: a gravity wave inner region of the radiative envelope, and a thermal wave region in the outer envelope. We show in Appendix \ref{app:dispersion1} and \ref{app:dispertion} that the transition occurs around a critical frequency $\omega_\mathrm{crit}\equiv (4\lambda N^2\omega_\mathrm{T})^{1/3}$, where $\omega_T=\kappa/r^2$ is the thermal frequency, and 
$\kappa = 16 \sigma_{\rm B} T^3/(3 \rho^2 c_{\rm P} \kappa_{\rm R})$ is the thermal diffusivity. The thermal mode exists where $|\omega_\mathrm{crit}|>\omega_\alpha$, while the g-mode exists where $|\omega_\mathrm{crit}|<\omega_\alpha$, as seen in the right panel of Figure \ref{fig:eigenfunction}. For higher frequency waves the thermal mode region becomes narrower and disappears. Since the thermal mode components only exist near the very surface of the star, where the density is very low, we would expect that the mechanical torques are mostly contributed by the travelling g-mode component excited in the deep interior as Zahn's theory assumed. Hence at low tidal forcing frequency the tidal torques should be similar to Zahn's model, as we will see in \S \ref{sec:results-wr}.

Figure \ref{fig:modes} shows the periods, damping rates and overlap integrals $Q_\alpha$ of all modes we solved for the same $10 \, M_\odot$ model. We see that most modes have a nearly constant period spacing, matching the expectations for g-modes. The damping rates for most modes are at the same order of magnitude, except at high-frequency where the damping is significantly lower. This is due to their low radial wave numbers $k_r$ and the damping rate $\gamma_\alpha\propto\int_\mathrm{star}k_r^2\kappa|\xi_\alpha|^2dm$. Low-frequency modes become traveling waves whose damping rate is roughly the wave crossing time.

The overlap integral $|Q_\alpha|$ typically decreases as the mode frequency decreases, but with significant scatter and with ``hills and valleys'' as the frequency decreases. Since the on-resonance AM dissipation $\dot{J}_\alpha\propto\gamma_\alpha|Q_\alpha|^2$, we expect to see the same ``hills and valleys'' features in the tidal synchronization rate, as the tidal forcing frequency is moving across different modes with varying $Q_\alpha$. This is also different from Zahn's theory, which predicts a ``smooth'' power-law relation for the AM deposition rate as a function of orbital period (\citealt{Kushnir2017}, or Equation \ref{eq:zahn_torque} in this paper).

In the frequency ranges where mixed modes appear, we notice that GYRE suffers from numerical convergence problems as it starts looking for extremely high-order modes. We identified some of the mode solutions in that regime as ``strange modes'', and an example is labelled in Figure \ref{fig:modes}. These modes often have excess damping rates and unusual winding numbers (mode radial order), and do not obey the usual frequency spacing of g-modes. In addition, their eigenfunctions appear to be artificially truncated as they reach deeper inside the star, unlike other modes with an inner g-mode component at similar frequencies, which are truncated near the convective core boundary. We are not sure if these modes are physical or caused by numerical artifacts from GYRE, hence we do not include them in the spin--orbit integration. A detailed discussion of these modes is presented in Appendix \ref{app:strange_modes}.

\section{Evolution of WR Spins}
\label{sec:results-wr}

We integrate the spin--orbit evolution of WR--BH binaries from the WR star models and oscillations modes we have computed. Throughout the evolution, the orbital AM of the system is lost due to winds from the primary, gravitational radiation and tidal AM transfer:
\beq
\dot{J}_\mathrm{orb}=\dot{J}_\mathrm{wind,orb}-\dot{J}_\mathrm{GW}-\dot{J}_\mathrm{tide}\,,
\eeq
where $\dot{J}_\mathrm{wind,orb}=\dot{M}_1\Omega_\mathrm{orb}(M_2a/(M_1+M_2))^2$ and $\dot{J}_\mathrm{tide}$ is given by summing over all modes from Equation \ref{eq:j_dissip}. At short orbits, the orbital decay timescale by gravitational wave radiation is given by \cite{Peters1964} (assuming circular orbits)

\beq
t_\mathrm{GW}\equiv\frac{a}{|\langle da/dt\rangle|}=\frac{5}{64(4\pi^2)^{4/3}}\frac{c^5(1+q)^{1/3}}{G^{5/3}M_1^{5/3}q}P_\mathrm{orb}^{8/3}\,,
\eeq
where $q=M_2/M_1$ is the mass ratio. This gives $t_\mathrm{GW}\approx206\,\mathrm{Myr}\times(M_1/10\,M_\odot)^{-5/3}(P_\mathrm{orb}/0.3\,\mathrm{d})^{8/3}$ for equal mass binaries ($q=1$), much greater than the typical WR lifetime ($\lesssim$ 1 Myr). Hence, gravitational radiation is not important in our case, but we still include the term $\dot{J}_\mathrm{GW}=(32/5)(G/a)^{7/2} c^{-5} M_1^2M_2^2\sqrt{M_1+M_2}$ in our evolution.

The primary receives spin AM from the orbit at $\dot{J}_\mathrm{tide}$, and it loses AM through winds:
\beq
\dot{J}_\mathrm{spin} = \dot{J}_\mathrm{tide}+\dot{J}_\mathrm{wind,spin}\,,
\eeq
where $\dot{J}_\mathrm{wind,spin}=\dot{M}_1\Omega_\mathrm{spin}R_1^2$. The spin of the primary may also change due to the changes of its internal structure and hence moment of inertia. Since the secondary is a BH in our case, its spin is not coupled. 

For comparison, we also integrate each evolution based on Zahn's formalism, with an adapted AM transfer rate from \cite{Kushnir2017}:
\beq
\label{eq:zahn_torque}
\dot{J}_\mathrm{tide,Zahn}=\frac{GM_2^2}{r_\mathrm{c}}\bigg(\frac{r_\mathrm{c}}{a}\bigg)^6s_\mathrm{c}^{8/3}\frac{\rho_\mathrm{c}}{\bar{\rho}_\mathrm{c}}\bigg(1-\frac{\rho_\mathrm{c}}{\bar{\rho}_\mathrm{c}}\bigg)^2\,,
\eeq
where $s_\mathrm{c}=\sqrt{3/\pi G\bar{\rho}_\mathrm{c}}|\Omega_\mathrm{orb}-\Omega_\mathrm{spin}|$, while $r_\mathrm{c},\rho_\mathrm{c}$ and $\bar{\rho}_\mathrm{c}$ are the convective core radius, the density at the core boundary, and the average density of the core, respectively.

We construct the integration machinery of the spin--orbit evolution as follows: after generating a grid of stellar model snapshots throughout the star's evolution, we solve for oscillation modes for each snapshot with GYRE. We begin our integration at the start of the helium burning phase (defined by the instant when 2\% of the core helium burning lifetime has passed). We carefully apply an adaptive time step control to avoid i) sudden crossing of resonance locations; ii) sudden changes of mode frequencies; iii) changes of more than 2\% of the total evolution phase lifetime; iv) sudden change of stellar spin by 2\%, in one time step. To evaluate physical quantities (e.g. mode frequencies, stellar masses) between two model snapshots, we estimate them by interpolating these snapshots and their corresponding GYRE solutions. In doing so, we track the modes by their radial orders $n_\mathrm{pg}$, and only include the mode eigenfunctions existing in both snapshots. We carried out resolution tests with half our selected timesteps and we confirm that the results are nearly identical.

Theories have suggested that the strong magnetic coupling between the stellar core and envelope (e.g. Taylor-Spruit dynamo, \citealt{Spruit2002, Fuller2019}) removes the majority of core AM immediately after the main sequence \citep{ma:19}, before the envelope can be stripped off. Hence, we assume initially non-rotating WR stars. We run models with initial orbital periods of 0.3, 0.5, or 0.8 days. We find that longer period orbits exhibit very little tidal spin-up. 

In this work we specify our calculations to equal mass binaries ($q=1$), since they are the most relevant for binary black holes. For cases with different mass ratios, one would expect from Equation \ref{eq:tide_dissip} that the tidal dissipation rate (hence the tidal spin-up rate) na\"ively scales as $q^2$, as we verified in some additional test runs. However, extreme mass ratios could allow for orbital decay and resulting processes such as resonance locking or binary mergers. We hope to generalize our calculations to such systems in future works.

\begin{figure*}
    \centering
    \includegraphics[width=\textwidth]{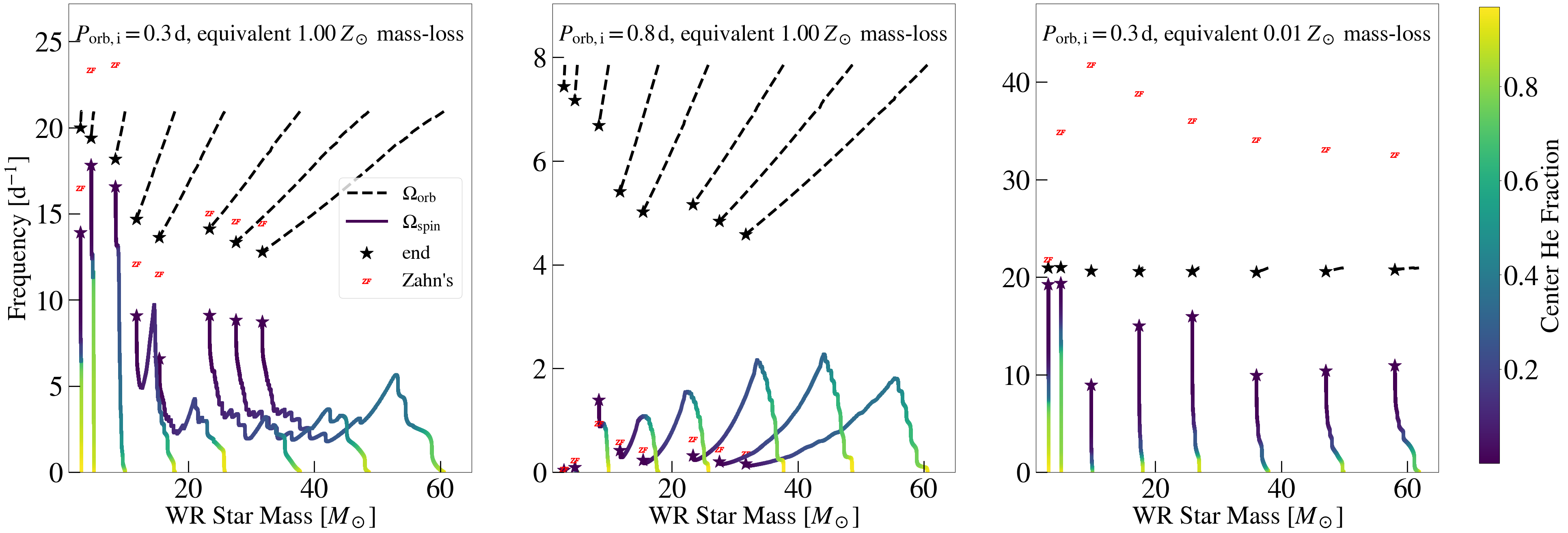}
    \caption{The spin and orbital evolution for our Wolf--Rayet--BH binaries. All systems have equal mass companions initially. The solid and dashed lines show the spin and orbital frequencies, respectively.  Line colors indicate the evolving central helium mass fraction. The stars mark the end of evolution (core helium depletion), and the red ``ZF'' symbols show the spins if Zahn's formalism (Equation \ref{eq:zahn_torque}) is assumed. {\bf Left}: Systems with initial orbital periods of 0.3 days and a mass-loss rate equivalent to solar metallicity. Mass loss is very significant for high-mass models and the final spins depart from Zahn's formalism for them. {\bf Middle}: Systems with initial orbital periods of 0.8 days and a mass-loss rate equivalent to solar metallicity. Mass loss overpowers tidal spin-up and the primaries are not spun up much, consistent with Zahn's results. {\bf Right}: Systems with initial orbital periods of 0.3 days and a mass-loss rate equivalent to 0.01 solar metallicity. Mass loss is negligible for most systems and the primaries are partially spun up, but not as much as Zahn's formalism predicts. None of these models reach tidal synchronization.}
    \label{fig:evo}
\end{figure*}

\begin{figure}
    \centering   \includegraphics[width=\columnwidth]{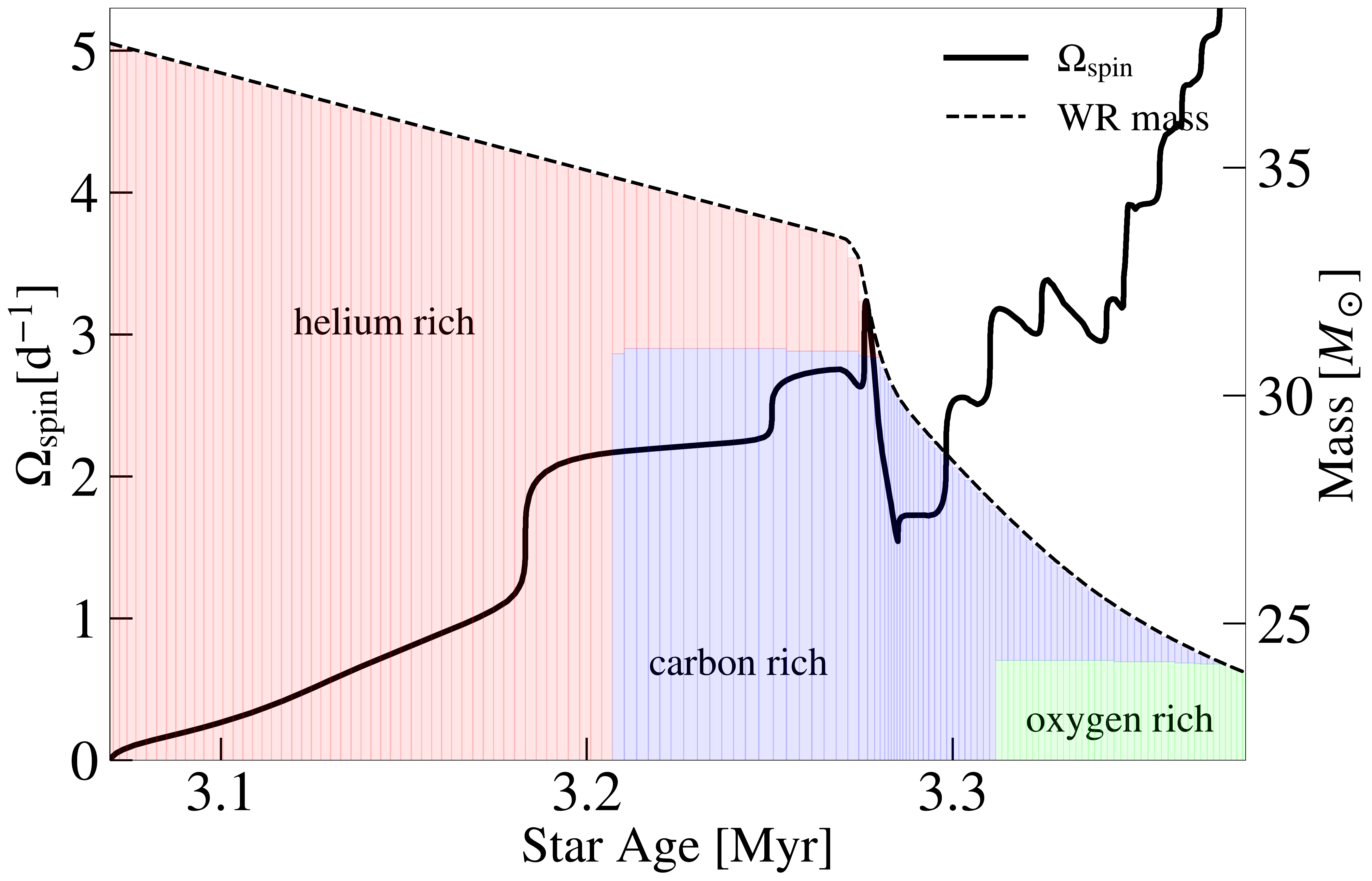}
    \caption{The mass and spin evolution of a $38\,M_\odot$ Wolf--Rayet star at solar metallicity, with an initial orbit of 0.3 days. The shaded regions show the dominant composition as a function of mass coordinate (right axis). At an age of $\sim3.3\,\mathrm{Myr}$, mass loss exposes the carbon-rich core, greatly enhancing the mass loss and spin down rates.}
    \label{fig:mass_loss}
\end{figure}

\begin{figure}
    \centering
    \includegraphics[width=0.925\columnwidth]{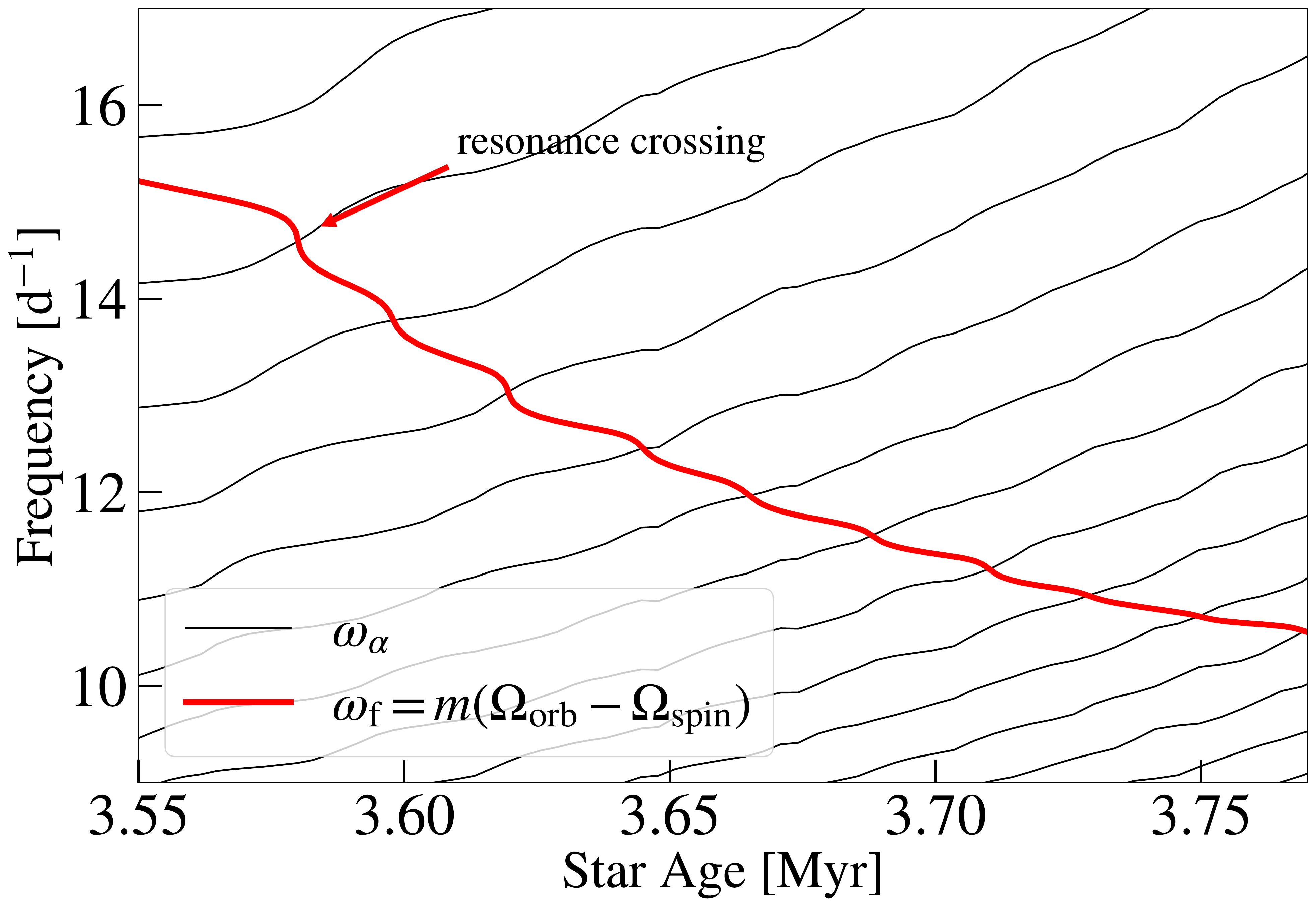}
    \caption{The evolution of the mode frequencies (black lines) and tidal forcing frequency (red line) of a WR--BH binary with a $26\,M_\odot,\,1.0\,Z_\odot$ WR star, an equal mass companion and an initial orbit of 0.8 days. The mode frequencies increase as the star evolves, while the forcing frequency decreases, preventing resonance locking. $\Omega_{\rm spin}$ increases rapidly while $\omega_\mathrm{f}$ decreases rapidly at resonance crossings, creating the ``step-like'' features we see in the spin frequency evolution (Figure \ref{fig:evo}).}
    \label{fig:omegas}
\end{figure}

Figure \ref{fig:evo} shows the spin and orbital evolution for a few of the systems we studied. Since the WR primary burns helium and loses mass throughout the evolution, the mass and central helium fraction can be seen as time coordinates, as shown. For systems with short initial orbits and high metallicities ($P_\mathrm{orb,i}=0.3\,\mathrm{d},\;Z=Z_\odot$, left panel), we see that the primaries get significantly spun up, yet they are not tidally synchronized even at the end of evolution. In addition, the final spins of massive models are never higher than what one would expect from Zahn's theory.

When we increase the initial periods ($P_\mathrm{orb,i}=0.8\,\mathrm{d}$, middle panel of Figure \ref{fig:evo}), the tidal torques decrease as expected. At these long periods, the tidal torque is dominated by traveling waves and the results agree well with Zahn's formalism. However, the tidal torque is unable to compete with mass loss, which almost completely removes the spin AM the star accumulated during the first half of the evolution, leaving a slowly spinning primary.

Tidal spin-up is followed by mass-loss induced spin-down in the middle of the evolution for solar-metallicity models (Figure \ref{fig:evo}, left and middle panel) due to an increase in the wind mass loss rate.
This occurs when the helium envelope is lost completely, exposing the CO-rich core, and greatly increasing the mass loss rate according to the ``Dutch" mass loss prescription (Figure \ref{fig:mass_loss}). In several thousand years the winds remove the star's spin AM until the mass loss rate decreases somewhat, allowing tidal spin-up to proceed. However, ongoing mass loss and a widened orbit prevent tides from spinning up the star to synchronization.

When we consider short initial periods but move to low-metallicity models ($P_\mathrm{orb,i}=0.3\,\mathrm{d},\;Z=0.01\,Z_\odot$, right panel of Figure \ref{fig:evo}), the mass loss becomes negligible and the spin evolution is dominated by tidal effects. The orbits do not change significantly. We see that the primaries get significantly spun up, yet still not reaching tidal synchronization, and the resulting spin is much slower than Zahn's prediction, except for the lowest mass models. This is because the transition period from standing modes to travelling waves increases as the stellar mass increases. Hence the evolution is more likely to depart from Zahn's formalism for more massive primaries (see \S \ref{sec:compare_zahn}).

The evolution of spin frequencies show ``step-like'' features (most easily seen in Figure \ref{fig:evo} middle panel) characterized by sudden increases in spin frequency. This is caused by the resonance-crossing of standing modes with the tidal forcing, as illustrated in Figure \ref{fig:omegas}. When the tidal forcing frequency gets close to one of the mode frequencies, a near-resonance occurs (cf. Equation \ref{eq:tide_dissip}) and the tidal torque drastically increases, leading to high spin-up rate. The occasional crossings of these resonances create the ``step-like'' features.

\section{Discussion}
\label{sec:discussion}

\begin{figure*}
    \centering
    \includegraphics[width=\textwidth]{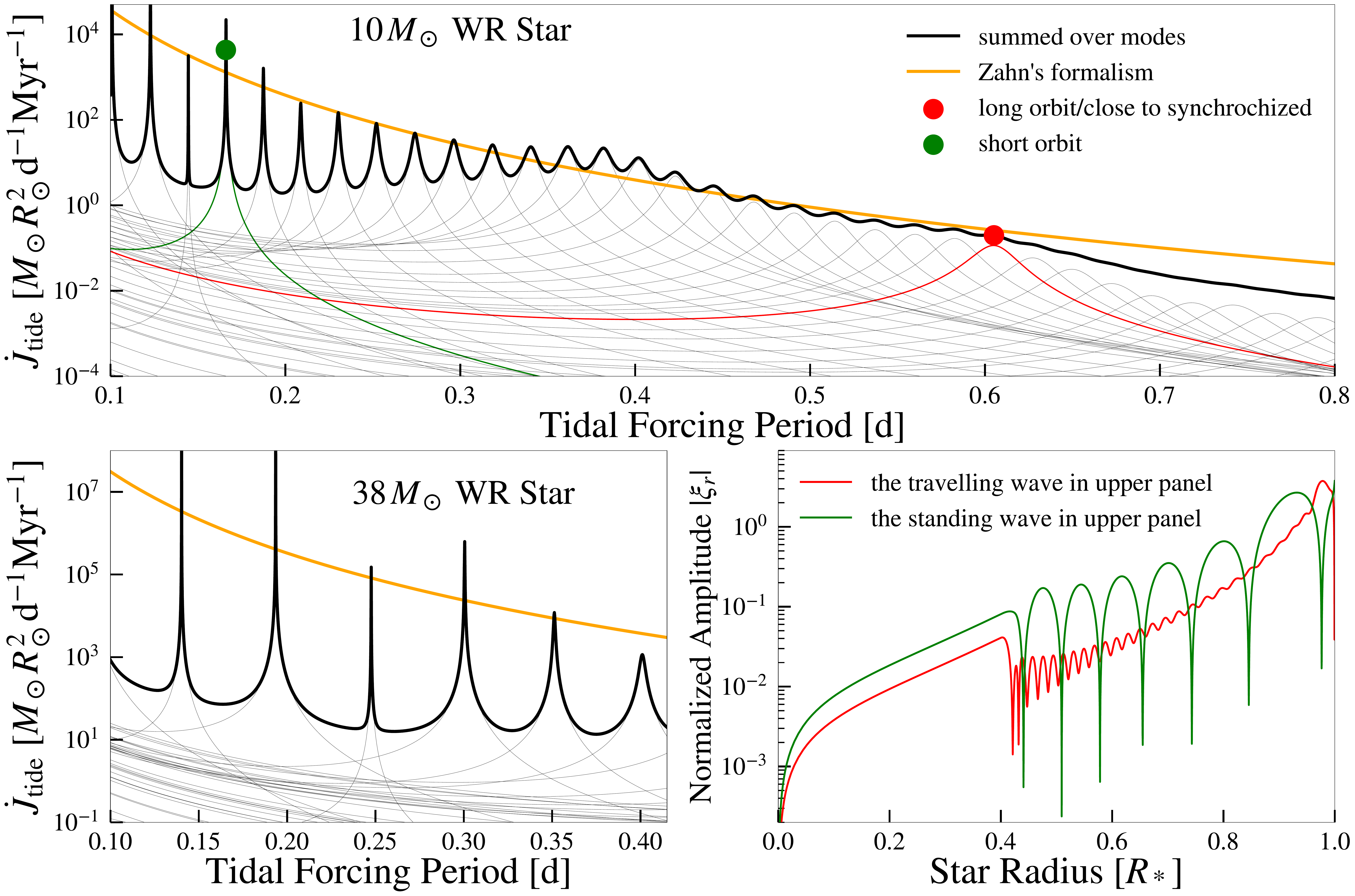}
    \caption{{\bf Upper}: Our calculated $\dot{J}_\mathrm{tide}$ by summing over modes (thick black line) and by Zahn's formalism (thick orange line) as a function of tidal forcing period $P_\mathrm{tide}=2\pi/\Omega_\mathrm{tide}$, for a $10\,M_\odot$ Wolf--Rayet model with $90\%$ central helium abundance. The contributions from individual modes are represented by thin gray lines. At short tidal forcing periods $P_\mathrm{tide}\lesssim 0.35\,\mathrm{d}$ (e.g., the green dot), the tidal torque is dominated by resonance peaks from individual standing modes, and is very different from what Zahn predicts. At long forcing periods (e.g., the red dot), the tidal torque is no longer dominated by individual modes, but arises from a multitude of highly-damped travelling modes. This is exactly Zahn's assumption, and hence the torques are similar at long periods. {\bf Lower Left}: The same as the upper panel but for a more massive $38\,M_\odot$ Wolf--Rayet model with $90\%$ central helium abundance. We see that the mode-period spacing becomes larger, and the standing wave region extends to longer tidal forcing period. {\bf Lower Right}: Eigenfunctions of the most resonant mode at the green and red dots in the upper panel. We see clearly that one is a standing mode while the other is a travelling mode.}
    \label{fig:compare_zahn}
\end{figure*}

\subsection{Comparison to Zahn's Formalism}
\label{sec:compare_zahn}

To understand why and how the tidal evolution differs from Zahn's formalism, in the upper panel of Figure \ref{fig:compare_zahn} we show the tidal torques calculated for a $10\,M_\odot$ WR star model for different tidal forcing periods $P_\mathrm{tide}=2\pi/\Omega_\mathrm{tide}$ with these two approaches. For short tidal periods, the tidal torque has sharp peaks at standing g-mode frequencies, in contrast to the power-law dependence of Zahn's prediction. This is caused by low damping rates of standing modes at short mode periods (cf. Figure \ref{fig:modes} and Figure \ref{fig:compare_zahn} lower right panel).

As the mode and orbital frequencies evolve, resonance crossings occur, hence the accumulated tidal spin-up must be evaluated by integrating $\dot{J}_\mathrm{tide}$ over time. Zahn's theory for travelling waves generally overestimates the tidal spin-up in this case, as the resonance peaks are narrow, and during the majority of evolution, the tidal torque is much less than what Zahn's formalism estimates. This helps to explain the departure of final spins shown in Figure \ref{fig:evo} for short-period systems (left and right panel) from Zahn's predictions.

When the systems are in long-period orbits, or already at a stage where $\Omega_\mathrm{spin}$ becomes comparable to $\Omega_\mathrm{orb}$ (close to tidal synchronization), the tidal forcing periods become long and the tidal torques are mostly contributed by modes at low frequencies. These modes, in contrast to their high-frequency partners, have large damping and are essentially travelling waves (cf. Figure \ref{fig:modes} and Figure \ref{fig:compare_zahn} lower right panel). Therefore, the tidal torque is no longer dominated by resonance with an individual mode, but instead has contributions from many strongly damped modes, effectively forming a ``continuum'' (Figure \ref{fig:compare_zahn}, red dot). This continuum formed by traveling waves is exactly what Zahn's formalism assumes, so the torques should be similar to Zahn's formalism, which is confirmed in Figure \ref{fig:compare_zahn}. This explains the consistency between our results and Zahn's for long-period initial orbits (Figure \ref{fig:evo} middle panel).

We note, however, that the transition period between standing waves and travelling waves depends on the stellar mass: for higher mass models, the frequency range for standing waves extends to longer periods, as shown for the $38\,M_\odot$ model in the lower left panel of Figure \ref{fig:compare_zahn}. Hence, the tidal evolution of massive WR stars departs more strongly from Zahn's formalism, as we see in the left and right panels of Figure \ref{fig:evo}.

\begin{figure}
    \centering
    \includegraphics[width=\columnwidth]{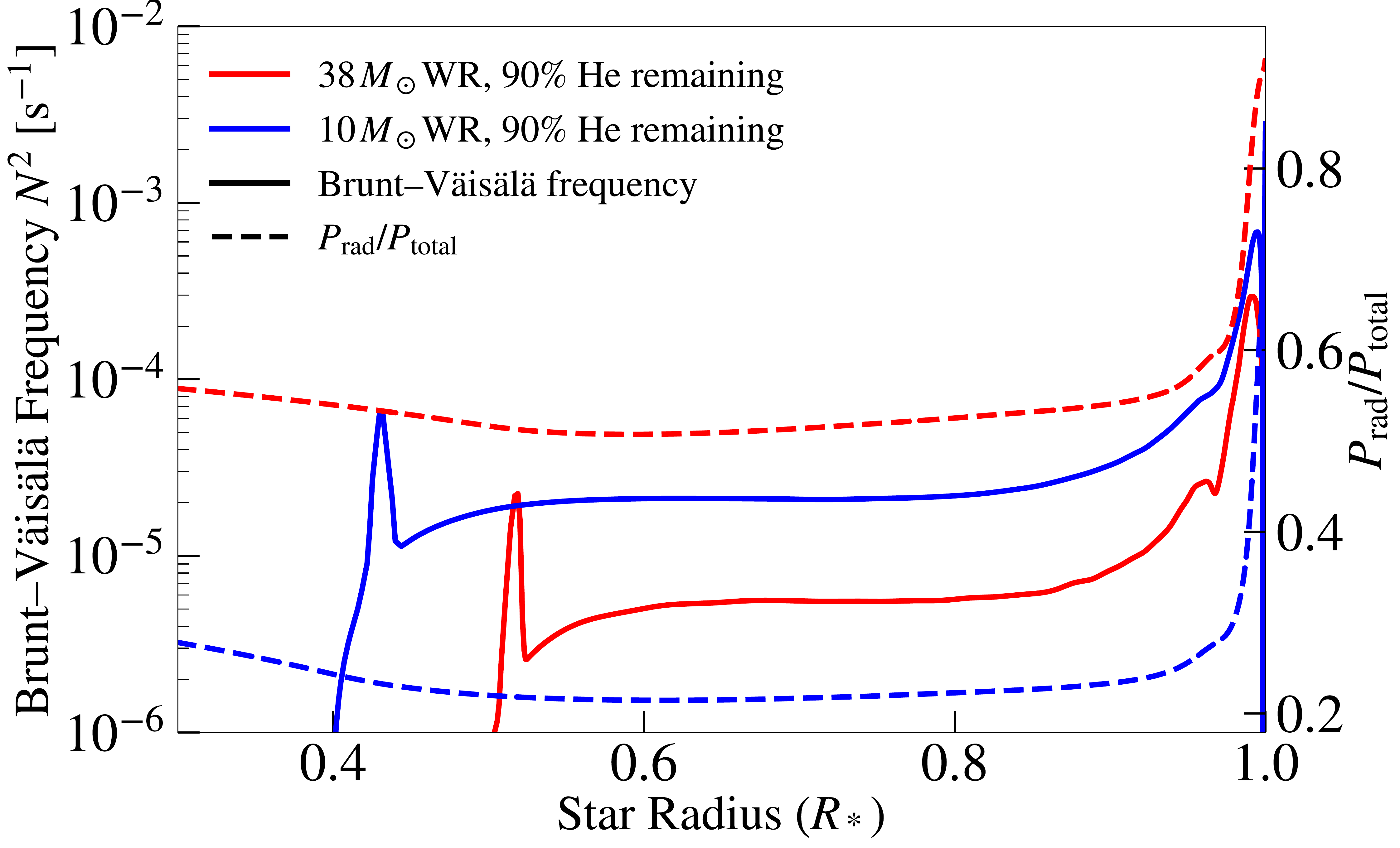}
    \caption{The Brunt-V\"ais\"al\"a frequency (solid lines) and the ratio of radiation pressure to total pressure (dashed lines) of two WR models of $10$ (blue) and $38\,M_\odot$ (red), the same models in Figure \ref{fig:compare_zahn}. The radiation pressure fraction is higher for the more massive model, making its Brunt-V\"ais\"al\"a frequency lower. This causes its resonance peaks to be narrower and more separated, as seen in Figure \ref{fig:compare_zahn}.}
    \label{fig:brunt}
\end{figure}

This distinction is caused by the different structures of low and high-mass WR stars. For higher mass stars, a larger fraction of the total internal pressure is contributed by radiation pressure since they are hotter and more luminous. Radiation pressure, however, contributes smaller buoyancy forces because the Brunt-V\"ais\"al\"a frequency $N$ is zero for a star supported purely by radiation pressure. Indeed, the Brunt-V\"ais\"al\"a frequencies within our high-mass models are smaller than those within our low-mass models (Figure \ref{fig:brunt}). This increases the g-mode period spacing (proportional to $N^{-1}$) of high-mass stars and decreases the radial wave number at the same frequency (as $k_r\propto N$). Hence, the mode damping rate $\gamma_\alpha\propto\int_\mathrm{star}k_r^2\kappa|\xi_\alpha|^2dm$ is also decreased. A secondary effect is that the convective cores are larger in more massive stars, making the radiative envelopes and g-mode cavities narrower. These combined effects make the resonance peaks narrower and more widely spaced for higher mass models, and further from the travelling wave limit of Zahn's formalism.

\subsection{Resonance Locking}

When the tidal dissipation is dominated by resonant modes, an important process called resonance locking may occur \citep{witte:99}. However, we argue that this is unlikely to occur for WR--BH binaries. Resonance locking can happen when the mode's frequency evolves at the same rate as the tidal forcing frequency:
\beq
\dot{\omega}_\alpha=\dot{\omega}_\mathrm{f}=m(\dot{\Omega}_\mathrm{orb}-\dot{\Omega}_\mathrm{spin})\,.
\eeq
We see from our example evolution tracks (Figure \ref{fig:evo}) that in most cases $\dot{\Omega}_\mathrm{orb}<0$ and $\dot{\Omega}_\mathrm{spin}>0$, which means $\dot{\omega}_\mathrm{f}<0$, in contrast to the fact that $\dot{\omega}_\alpha>0$ due to stellar evolution (Figure \ref{fig:omegas}). Hence the above relation never holds and resonance locking can never happen. Instead, the system rapidly passes through resonances, creating the step-like features in Figure \ref{fig:evo}. 

We note that when mass loss dominates the spin evolution, we could occasionally have $\dot{\Omega}_\mathrm{spin}<0$ (Figure \ref{fig:evo}, left and middle panels) and resonance locking may happen during this phase. This may prevent a star from rapidly spinning down, but it cannot cause tidal spin-up. However we find that during these phases the Brunt-V\"as\"ail\"a frequency of the star increases rapidly, such that $\dot{\omega}_\alpha$ exceeds $\dot{\omega}_\mathrm{f}$ even at resonance, in contrast to the resonance locking criterion. Hence, resonance locking does not appear to occur in any of our modeled systems.

\subsection{Implications for BH Spins}

A rapidly rotating WR star can probably collapse to a fast-spinning BH, forming a high-spin binary BH system. If angular momentum is conserved during the core-collapse process, the dimensionless spin parameter of the resulting BH is
\beq
a=\frac{cJ_\mathrm{WR}}{GM_\mathrm{BH}^2}\,.
\eeq
In Figure \ref{fig:bh_spin}, we show the resulting black-hole spins of our WR star models, assuming that they preserve their masses and angular momenta after helium burning and during the core-collapse process. We also show the predicted BH spins with Zahn's formalism. We see that lower-mass systems can form faster-spinning BHs, as their tidal spin-up is more efficient. It is only in ultra-short orbits that these systems form fast-spinning BHs. For solar metallicity systems, tidal spin-up cannot overcome AM loss from winds, resulting in low spins for systems starting at long ($P_\mathrm{orb,i}\gtrsim 0.5\,\mathrm{d}$) orbital periods.  Low-metallicity ($1\%\,Z_\odot$) systems with $0.5 \, {\rm d} \! \lesssim \! P_\mathrm{orb,i} \! \lesssim \! 1\,\mathrm{d}$ produce larger BH spins with $0.1\lesssim a\lesssim 0.8$, compared to $a\sim0.01$ predicted by single star evolution models \citep{fuller:19}. For high-mass systems, the spins are much smaller than Zahn's predictions. 

If the companion black hole is assumed non-spinning, our predicted BH spins will be roughly compatible with some LIGO measurements with moderate spins ($0.1\lesssim \chi_\mathrm{eff}\lesssim 0.5$) \citep{Abbott2019,Abbott2019PRX,Abbott2021}, but would have a tough time matching any events with large $\chi_{\rm eff}$. The relationship between orbital period, mass, and spin is different than what Zahn's theory predicts. Whereas we typically find higher spins for $M_{\rm BH} \lesssim 10 \, M_\odot$, Zahn's theory predicts smaller spins for lower mass BHs. There may be an anti-correlation between mass and spin \citep{safarzadeh:20} which would support our new models. A mass-spin correlation from future LIGO-VIRGO data will help distinguish between these models.
None of our high-mass models predict spins comparable to some high-spin measurements ($a\gtrsim 0.9$) from X-ray binaries \citep{Miller2015}, and the uncertainty of such measurements is still under debate \citep{Belczynski2021,Fishbach2022}. However, those measurements are for the first-born BH, while our models only apply to the second-born BH.

Previous works on tidal interactions between WR--BH binaries have predicted black hole spins similar to our ``Zahn's results'' in Figure \ref{fig:bh_spin}, in which Zahn's formalism is assumed \citep{Qin2018, Bavera2020, Belczynski2020, Olejak2021, fuller2022}. These results likely overestimate the black hole spins when standing waves are present, which applies primarily to massive BHs ($M_{\rm BH} \gtrsim 10 \, M_\odot$). \citealt{Detmers2008} also investigated tidal spin-up of WR stars, but used different prescriptions for tidal dissipation, winds, and orbital AM losses. Unlike our results and those listed above, they found that tidal spin-up coupled with mass loss frequently caused the orbits to decay and instigate mass transfer. This outcome is more likely with small companion masses (e.g., neutron stars) whose orbits must decay more in order to tidally spin-up the WR star.

\label{sec:results-bh}

\subsection{Nonlinear dissipation}

Throughout the paper, we have assumed that the tidally forced modes are linear. However, this is not true when a mode is close to resonance, especially for massive models with larger on-resonance mode amplitudes. To examine how nonlinearity may affect our conclusions, we estimate the nonlinear damping rate for a mode $\alpha$ in Appendix \ref{app:nonlinear}. For weakly nonlinear modes, an approximate nonlinear damping rate may be
\beq
\gamma_{\alpha,{\rm NL}} \sim \frac{(d\xi_{\alpha,r}/dr)_\mathrm{max}}{\tau_{\alpha,2}} \, ,
\eeq
where the numerator is the mode nonlinearity (i.e., the peak value of $d\xi_r/dr$ within the star, which is much less than unity for a weakly nonlinear mode) and the denominator is the wave crossing time of the envelope. We rerun our evolution models with this nonlinear damping rate, and we find that nearly all models achieve more tidal spin-up compared to linear damping only. 

\begin{figure}
    \centering
    \includegraphics[width=\columnwidth]{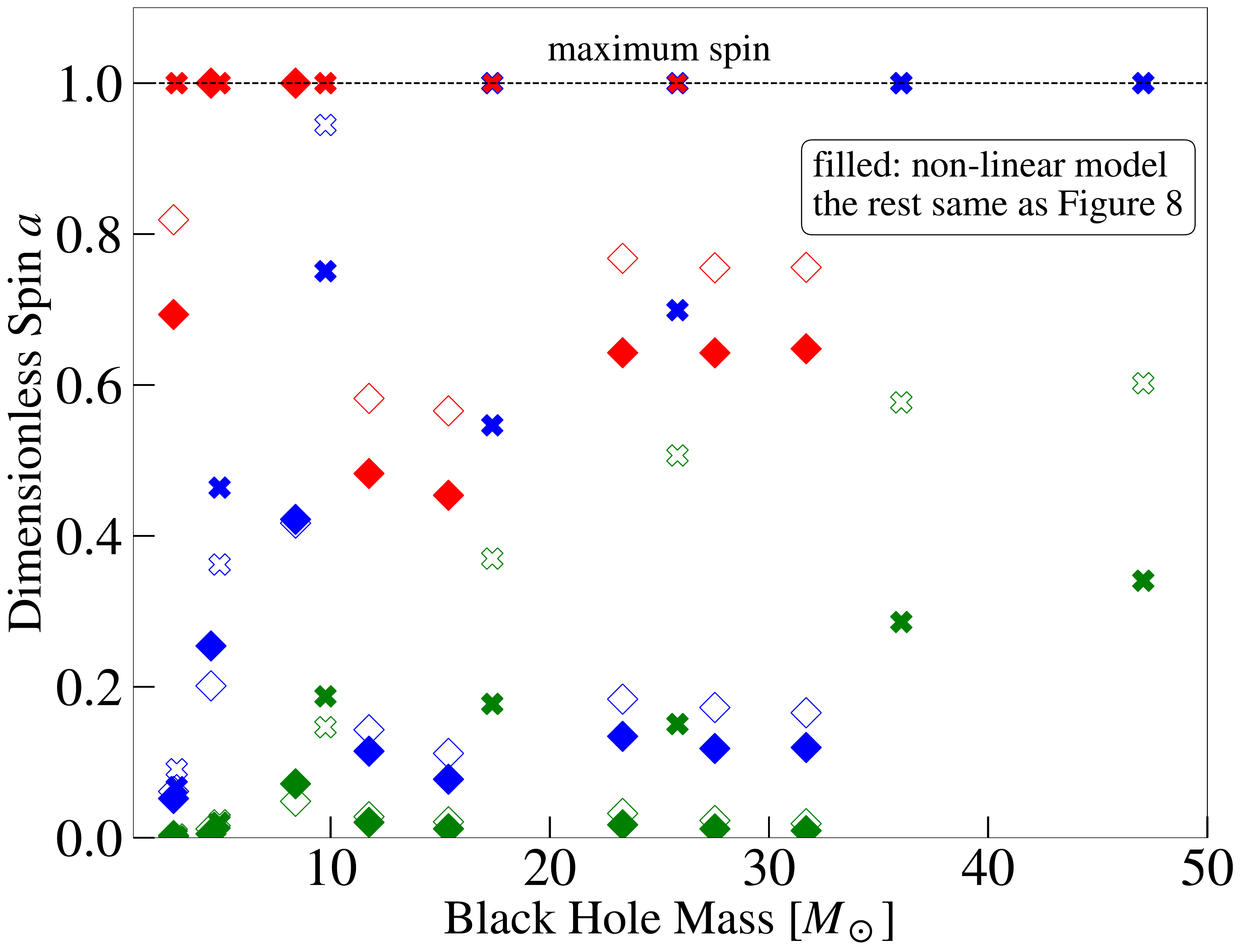}
    \caption{The dimensionless spin of resulting black holes for our Wolf--Rayet star models with the nonlinear damping rates we estimate in Appendix \ref{app:nonlinear}. Zahn's predictions are also shown. The black hole spins are significantly increased when including nonlinear damping, especially for high-mass black holes and short-period systems. Hence, nonlinear effects could be an important factor in these systems.}
    \label{fig:bh_spin_nl}
\end{figure}

In Figure \ref{fig:bh_spin_nl}, we show the predicted BH spins for our systems with nonlinear damping, compared to the predictions from Zahn's theory. We see that for very short-period orbits ($P\sim0.3\,
\mathrm{d}$), the strong tidal forcing triggers substantial nonlinear damping, spinning up the BHs to nearly the same rotation rates as predicted by Zahn's theory, where maximum damping occurs. Nonlinear effects are the most significant for low-metallicity and high-mass ($M\gtrsim 30\,M_\odot$) models, which have lower order g-modes dominating their tidal processes and less linear damping (see discussion in \S \ref{sec:compare_zahn}).

However, our nonlinear damping model is crude, so these predictions are not very reliable. A more detailed study of the nonlinear interactions has to be carried out to establish firm conclusions for the final BH spins.

\subsection{Caveats}
\label{angmom}

Throughout this paper, we have assumed very efficient angular momentum transport within the WR star, such that it remains rigidly rotating. This is justified by the asteroseismically callibrated models of magnetic angular momentum transport \citep{fuller:19} that predict nearly rigid rotation during the helium-burning phase \citep{fuller2022}. However, if angular momentum transport is inefficient, gravity waves damping near the stellar surface will preferentially spin up those layers until they are synchronized. This will create a critical layer at which subsequent waves are absorbed \citep{goldreich:89}, synchronizing the star from the outside inwards. Recent works have investigated the formation of critical layers and subsequent absoroption of incoming waves \citep{su:20,guo:22}, though they do not include magnetic torques that may allow angular momentum transport to prevent critical layer formation. If a critical layer can form, it will absorb outgoing waves, such that Zahn's model applies once again.

We have ignored the influence of the Coriolis force in our calculations. This will become significant once the star's have been partially spun up and the tidal forcing frequency becomes smaller than the rotation frequency. However, the prograde $\ell=m=2$ modes that dominate the tidal interaction have eigenfunctions that are only slightly changed by Coriolis forces (see, e.g., \citealt{fuller:14}), so we don't expect any of our conclusions to be greatly affected.

\begin{figure}
    \centering
    \includegraphics[width=\columnwidth]{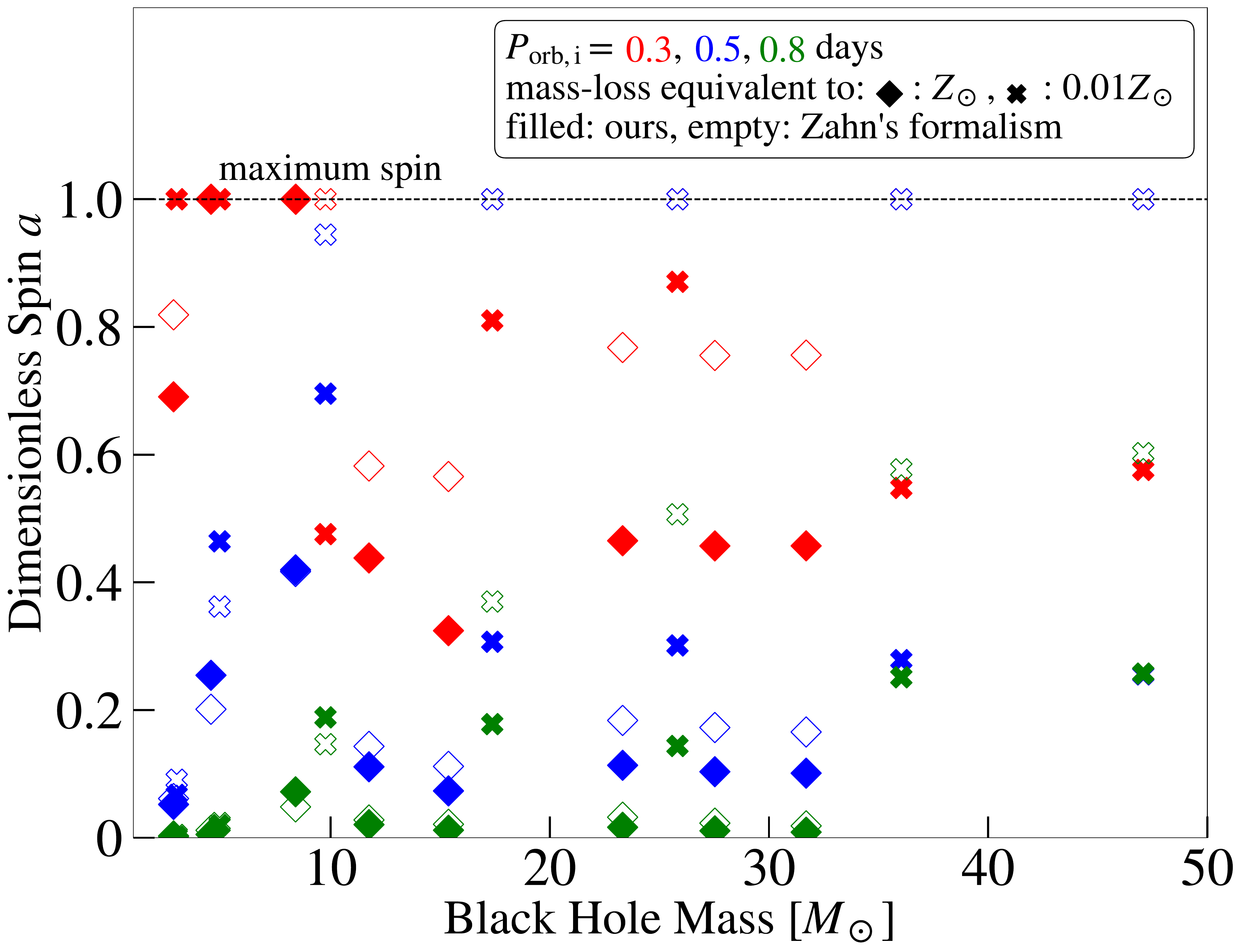}
    \caption{The dimensionless spin of resulting black holes for our Wolf--Rayet star models if their masses and angular momenta at the end of helium burning is preserved. Zahn's predictions for the same system are also shown, and are assumed 1 (maximum rotating) if the progenitors have $J>GM^2/c$. For short initial orbits, the models typically predict higher spins than individual stellar evolution models, where the spins could be as low as $10^{-2}$. However, the spins are usually lower than Zahn's predictions, especially for high-mass systems.}
    \label{fig:bh_spin}
\end{figure}

We have adopted the ``Dutch'' wind models with an artificial scaling factor to simulate the mass-loss rates for Wolf--Rayet stars at different metallicities. However, the wind physics for these stripped stars are highly uncertain \citep{Vink2022}, and different wind models could result different rates for the removal of spin angular momentum from the primary, introducing uncertainties in the final black hole spins. Nevertheless, we don't expect these uncertainties to exceed the differences between our solar-metallicity models and the $0.01\,Z_\odot$ models, as they represent extreme cases of large and negligible mass-loss, respectively. Hence, the final black hole spins with the ``correct'' wind physics should lie between the data points representing models with the same initial mass and periods but different metallicities in our Figure \ref{fig:bh_spin}. The conclusion that these black holes are not spun up to maximal rotation appears robust.

Our stellar models were run at low metallicity in order to reliably calculate the near-surface structure and mode eigenfunctions. Higher metallicity stars will have somewhat different structure and mode eigenfunctions near the surface, particularly around the iron group opacity peak. If this significantly affects mode damping rates, then the tidal synchronization efficiency would be similarly altered. We set up additional test models with $0.02,\,0.03$ and $0.2\,Z_\odot$ and find that the oscillation mode parameters (frequencies, damping rates and overlap integrals) show no significant differences, nor specific trends towards higher metallicities, hence we expect our treatment to be appropriate. However, these models all have weak winds, while the strong winds in solar-metallicity models may also alter the eigenmode properties. \cite{ro:16,ro:19} present detailed models of the transition from the hydrostatic star the hydrodynamic wind in the near-surface layers. Future work should investigate how those types of stellar models affect mode eigenfunctions and damping rates. 

Finally, our calculations are performed by summing up the contribution of individual tidally excited oscillation modes. If non-resonant modes outside our computed frequency range contribute to the tidal dissipation, or if our eigenmode calculations miss highly non-adiabatic thermal modes that contribute to the dissipation, then the tidal dissipation rate could possibly increase. It would be interesting to compare to calculations performed by directly computing the forced tidal response, as outlined in \citep{sun:23}.

\section{Conclusion}
\label{sec:conclusion}

In this work, we investigate the dynamical tidal spin-up of Wolf--Rayet stars from black hole companions. We build Wolf--Rayet star models with different metallicities, and then calculate their oscillation mode frequencies, damping rates, and eigenfunctions. We use these to integrate the coupled spin--orbit evolution of the binary based on the tidal excitation of these oscillation modes. We also make predictions for the resulting BH spins upon core-collapse of the Wolf--Rayet star.

We study the properties of the oscillation modes and find that at shorter orbital period, the tidal forcing is mostly contributed by standing g-modes, in contrast to the usual assumption of travelling waves proposed by \cite{zahn:75}. The standing g-mode spectra contributes a resonance structure, and during most of the spin--orbit evolution phase, the tidal response lies between resonances and the interaction strength is weaker than Zahn's prediction. The tidal forcing transits to Zahn's travelling wave limit at longer periods, in which Zahn's estimate is more accurate. However, the specific transition frequency depends on the stellar masses, and the structure for more massive stars (supported significantly by radiation pressure) tend to have lower transition frequencies, allowing systems in longer-period orbits to evolve differently compared to Zahn's prediction.

We find that it is difficult to tidally synchronize Wolf--Rayet stars during helium-burning. For solar-metallicity Wolf--Rayet stars, strong winds tend to remove the majority of angular momentum deposited by tides, leaving slowly spinning stars and black holes. At low metallicity, the stellar wind is weaker and the stars are significantly tidally spun up, yet still less than Zahn's prediction, especially for massive stars and short-period orbits.

Tidal interactions can significantly spin up the resulting BHs compared to single-star models. Yet the predicted black hole spins $a$ are still $\lesssim \! 0.4$ for all but our shortest period ($P_\mathrm{orb}\lesssim0.5\,\mathrm{d}$) models. These predictions are consistent with some low/moderate-spin measurements from LIGO/Virgo black hole merger events, but cannot explain high-spin X-ray binaries events since only the second-born black hole has large spin in these models.

We have discussed a new class of gravito-thermal modes that appear in our calculations, yet we do not reach a firm conclusion whether these modes are physical or caused by numerical artifacts. Future work should investigate the origins of these modes, and any effect they could have on tidal spin-up.

There are also caveats to our work. We have assumed rigid rotation of the Wolf--Rayet star during our spin--orbit evolution calculations, as expected if there are strong internal AM transport processes caused by magnetic torques. However, weak AM transport could enable surface critical layer formation, allowing Zahn's model to apply. We did not realistically calculate the near-surface structure of our solar-metallicity models, which could alter our estimate of the mode damping rates. We also point out that nonlinear damping effects could be significant for our most massive models, which can produce more tidal dissipation than our predictions from linear theory. This should be studied and improved in future work.

\begin{acknowledgments}
We thank the anonymous referee for a constructive report that helped to improve this work. We also thank Hang Yu, Rich Townsend, Emily Hu and Katie Breivik for helpful discussions. This work is partially supported by NASA through grant 20-XRP20 2-0147. J.F. is thankful for support through an Innovator Grant from The Rose Hills Foundation, and the Sloan Foundation through grant FG-2018-10515.
\end{acknowledgments}

\software{MESA \citep{paxton:11,paxton:13,paxton:15,paxton:18,paxton:19}, GYRE \citep{townsend:13,townsend2018,goldstein2020}}

\appendix

\section{Mode Dispersion Relation with Thermal Diffusion and Radiation Pressure}
\label{app:dispersion1}

To understand the effects of thermal diffusion on stellar oscillation modes, we modify the derivations in the appendices of \cite{ma:19}, assuming the stellar interior has a mixture of ideal gas and radiation pressure, and constant molecular weight. The internal energy density is given by
\beq
\label{eq:energy}
u=c_\mathrm{V,g}T+aT^4\,,
\eeq
where $T$ is the temperature of the fluid, and $c_\mathrm{V,g}=n k_\mathrm{B}/(\gamma_\mathrm{g}-1)$ is the heat capacity for gas at constant volume, and $a$ is the radiation constant, $n$ is the number density of gas particles and $\gamma_\mathrm{g}$ is the heat capacity ratio for ideal gas ($\gamma_\mathrm{g}=5/3$ for mono-atomic gas). The pressure of the mixture is given by
\beq
\label{eq:pressure}
P=P_\mathrm{gas}+P_\mathrm{rad}=nk_\mathrm{B}T+\frac13 aT^4\,.
\eeq
Now we consider a change in entropy: from the first law of thermodynamics, we have $dS=(dU+pdV)/T$. This immediately leads to the change in entropy density
\beq
\label{eq:entropy1}
ds=nk_\mathrm{B}\bigg[\bigg(\frac{1}{\gamma_\mathrm{g}-1}+\frac{12\eta}{1-\eta}\bigg)d\ln T-\frac{1+3\eta}{1-\eta}d\ln \rho\bigg]\,,
\eeq
where we defined $\eta\equiv P_\mathrm{rad}/P$ and used $d\ln\rho=-d\ln V$, where $\rho$ is the gas density. Taking the derivative of Equation \ref{eq:pressure}, we have the following relation:
\beq
\label{eq:derivative}
d\ln P=(1-\eta)d\ln \rho+(1+3\eta)d\ln T\,.
\eeq
We substitute this relation into Equation \ref{eq:entropy1} to have two alternative forms of the entropy derivative:
\beq
\label{eq:entropy2}
ds=nk_\mathrm{B}\bigg[\bigg(\frac{1}{\gamma_\mathrm{g}-1}+\frac{12\eta}{1-\eta}+\frac{(1+3\eta)^2}{(1-\eta)^2}\bigg)d\ln T-\frac{1+3\eta}{(1-\eta)^2}d\ln P\bigg]\,,
\eeq
and
\beq
\label{eq:entropy3}
ds=nk_\mathrm{B}\bigg[\bigg(\frac{1}{\gamma_\mathrm{g}-1}+\frac{12\eta}{1-\eta}\bigg)\frac{1}{1+3\eta}d\ln P-\bigg(\frac{1-\eta+12(\gamma_\mathrm{g}-1)\eta}{(\gamma_\mathrm{g}-1)(1+3\eta)}+\frac{1+3\eta}{1-\eta}\bigg)d\ln \rho\bigg]\,.
\eeq
From Equations \ref{eq:entropy2} and \ref{eq:entropy3} we can calculate the following thermodynamic quantities:
\beq
c_\mathrm{P}\equiv T\bigg(\frac{\partial s}{\partial T}\bigg)_P=\bigg(\frac{\partial s}{\partial \ln T}\bigg)_P = \bigg(\frac{1}{\gamma_\mathrm{g}-1}+\frac{12\eta}{1-\eta}+\frac{(1+3\eta)^2}{(1-\eta)^2}\bigg)nk_\mathrm{B}\,,
\eeq
\beq
\label{eq:Gamma_1}
\Gamma_1 \equiv\bigg(\frac{\partial \ln P}{\partial \ln \rho}\bigg)_s=1-\eta+\frac{(\gamma_\mathrm{g}-1)(1+3\eta)^2}{1-\eta+12(\gamma_\mathrm{g}-1)\eta}\,.
\eeq

We now derive the energy equation for the mixture of ideal gas and radiation. With thermal diffusion, the entropy changes at a rate
\beq
\frac{ds}{dt}=\frac{\partial s}{\partial t}+{\bf v}\cdot\nabla s=\frac{c_\mathrm{P}\kappa}{T}\nabla^2 T\,.
\eeq
where $\kappa$ is the thermal diffusivity. We assume a static and spherically symmetric stellar background and the usual harmonic time dependence of perturbations $\delta Q\propto e^{-\sigma t}=e^{-i\omega t}$. The above equation reduces to
\beq
\label{eq:energy1}
-\sigma\bigg(\delta s+\xi_r\frac{\partial s}{\partial r}\bigg)=-c_\mathrm{P}\kappa k^2\delta\ln T\,,
\eeq
where $\xi_r$ is the radial displacement and we used the WKB approximation $\nabla^2\rightarrow-k^2$. From Equations \ref{eq:entropy3} and \ref{eq:Gamma_1}, we have
\beq
\frac{\partial s}{\partial r}=\frac{1-\eta}{1+3\eta}c_\mathrm{P}\bigg(\frac{1}{\Gamma_1}\frac{\partial \ln P}{\partial r}-\frac{\partial \ln \rho}{\partial r}\bigg)=\frac{1-\eta}{1+3\eta}\frac{c_\mathrm{P}}{g}N^2\,,
\eeq
where $N^2\equiv g(\Gamma_1^{-1}\partial\ln P/\partial r-\partial\ln \rho/\partial r)$ is the Brunt-V\"ais\"al\"a frequency. We can further express $\delta s$ and $\delta \ln T$ in terms of $\delta P$ and $\delta \rho$ from Equations \ref{eq:entropy3} and \ref{eq:derivative}. We substitute them and the above equation into Equation \ref{eq:energy1}, and arrive at our energy equation:
\beq
\label{eq:energyeq}
\bigg(1-\frac{\kappa k^2}{\sigma}\bigg)\frac{\delta\rho}{\rho}=\bigg(\frac{1}{\Gamma_1}-\frac{\kappa k^2}{\sigma}\frac{1}{1-\eta}\bigg)\frac{\delta P}{P}+\frac{N^2}{g}\xi_r\,.
\eeq
It is straightforward to verify that Equation \ref{eq:energyeq} reduces to the energy equation in \cite{ma:19} when $\eta=0$, i.e. radiation is neglected.

We now consider the dynamics of the fluid. The perturbed momentum equation reads:
\beq
\rho\omega^2\xi_r = ik_r\delta P+g\delta \rho\,,\quad\rho\omega^2{\bf \xi}_\perp = \nabla_\perp\delta P\,,
\eeq
where we again used the WKB approximation $\nabla_r\rightarrow ik_r$. The continuity equation with the incompressible approximation\footnote{The result is similar without this approximation.} gives
\beq
\nabla\cdot{\bf \xi}\approx ik_r\xi_r+\nabla_\perp\cdot{\bf \xi}_\perp=0\,.
\eeq
When the angular dependence of perturbation variables are expanded in spherical harmonics, we have $\nabla_\perp^2\rightarrow-\lambda/r^2$ where $\lambda=l(l+1)$. Combining the above equations with the energy equation, we arrive at the dispersion relation
\beq
1-\frac{\kappa k^2}{\sigma}+\frac{\lambda}{k^2r^2}\frac{N^2}{\sigma^2}=\bigg(\frac{1}{\Gamma_1}-\frac{\kappa k^2}{\sigma}\frac{1}{1-\eta}\bigg)\frac{ik_r}{k^2H}\,,
\eeq
where $H\equiv P/(\rho g)$ is the pressure scale height. With the WKB approximation, $k\approx k_r$ and $k_rH\gg1$, the first term in the bracket of the right hand side can be neglected, and the dispersion relation becomes
\beq
\label{eq:disp}
1-\frac{\kappa k_r^2}{\sigma}\bigg(1-\frac{1}{1-\eta}\frac{i}{k_r H}\bigg)\approx-\frac{\lambda}{k_r^2r^2}\frac{N^2}{\sigma^2}\,.
\eeq

\section{Gravity and Thermal Mixed Modes}
\label{app:dispertion}

\begin{figure}
    \centering
    \includegraphics[width=0.5\columnwidth]{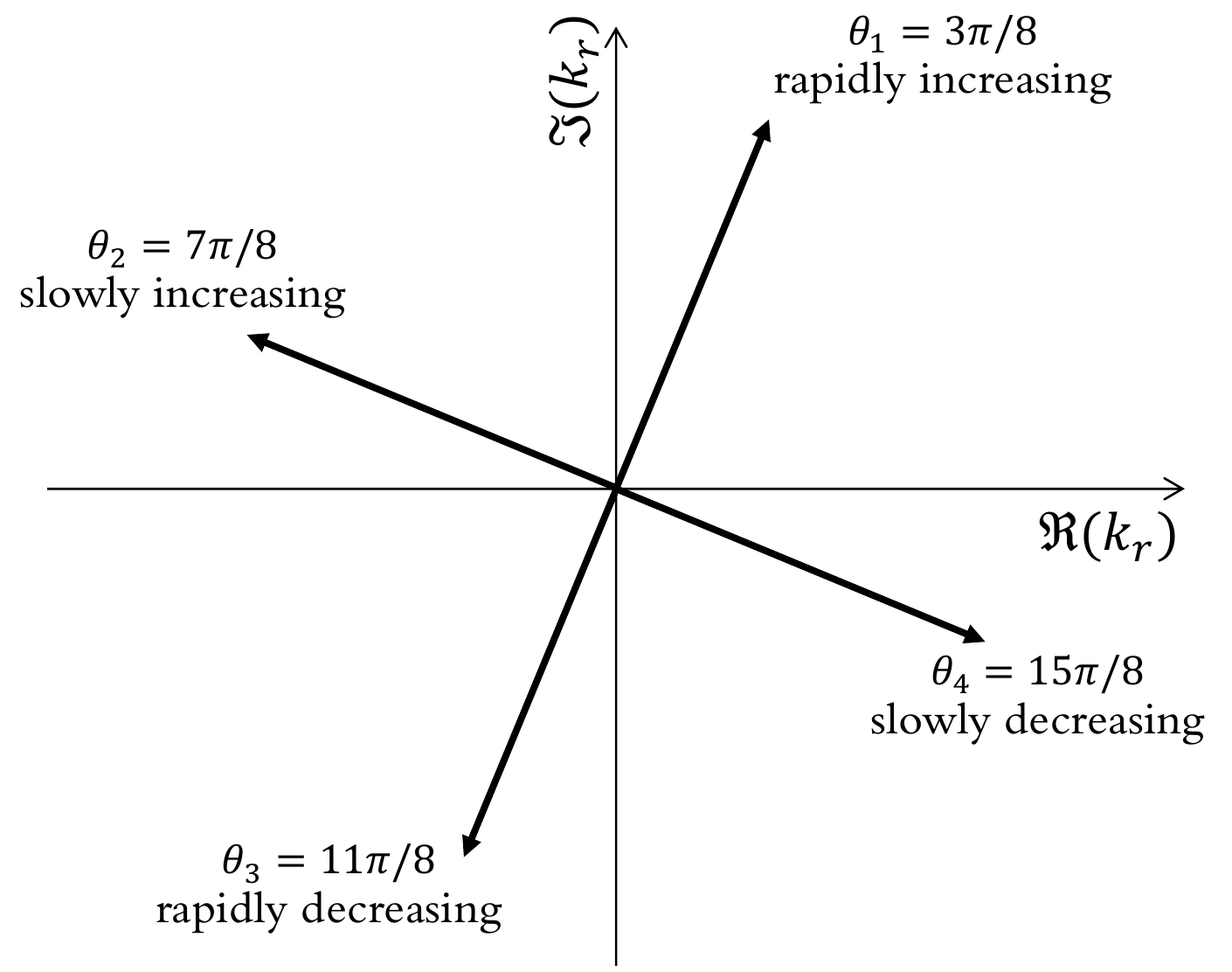}
    \caption{$k_r$ for the thermal mode on the complex plane. The four solutions correspond to rapidly increasing, slowly increasing, rapidly decreasing and slowly decreasing thermal modes, respectively.}
    \label{fig:k_complex_plane}
\end{figure}

When gas pressure is non-negligible, we always have $1-\eta\sim1$ and the second term in the bracket of the left hand side of Equation \ref{eq:disp} can usually be neglected under WKB approximation $k_rH\gg1$. The dispersion relation hence reduces to the quadratic equation
\beq
\label{eq:dispersion_quadratic}
\bigg(\frac{\omega_T}{\sigma}k_r^2 r^2\bigg)^2 - \bigg(\frac{\omega_T}{\sigma}k_r^2 r^2\bigg)-\frac{1}{4}\frac{\omega_\mathrm{crit}^3}{\sigma^3} = 0\,,
\eeq
where $\omega_T=\kappa/r^2$ is the thermal frequency, and 
\beq
\label{eq:omega_crit}
\omega_\mathrm{crit}\equiv (4\lambda N^2\omega_T)^{1/3}
\eeq
is the critical frequency between different types of modes. The solution to Equation \ref{eq:dispersion_quadratic} is
\beq
k_r^2r^2\frac{\omega_T}{\sigma}=\frac12\pm\frac12\bigg(1+\frac{\omega_\mathrm{crit}^3}{\sigma^3}\bigg)^{1/2}\,,
\eeq
which has two important limits.
\begin{enumerate}
    \item{\bf High-frequency region} ($|\omega_\mathrm{crit}|\ll|\sigma|$): the solution reduces to
    \beq
    k_r^2r^2\frac{\omega_T}{\sigma}\simeq\frac12\pm\frac12\bigg(1+\frac12\frac{\omega_\mathrm{crit}^3}{\sigma^3}\bigg)\,.
    \eeq
    The ``$+$'' sign solution further reduces to the (radial) thermal diffusion solution $k_r^2 \kappa \simeq \sigma$. The ``$-$'' sign solution reduces to $k_r^2=-\lambda N^2/r^2\sigma^2$, which is the g-mode dispersion relation. With $\sigma=i(\omega+i\gamma)$, under the weakly damped limit $\gamma\ll\omega$, we have
    \beq
    k_r\approx\pm\frac{\sqrt{\lambda} N}{r\omega^2}(\omega-i\gamma)\,,
    \eeq
    which means the wave amplitude increases/decreases as it gets closer to the envelope, at a rate $\Im(k_r)=\pm\sqrt{\lambda}N\gamma/(r\omega^2)$.
    
    \item{\bf Low-frequency region} ($|\omega_\mathrm{crit}|\gg|\sigma|$): The solutions reduce to
    \beq
    \label{eq:krtherm}
    k_r^2r^2\simeq
    \pm\bigg(\frac{\lambda N^2}{\omega_T\sigma}\bigg)^{1/2}\,.
    \eeq
    Under the weakly damped limit, $\sigma=i(\omega+i\gamma)\approx i\omega$, we hence have
    \beq
    k_r\approx\frac{e^{i\theta}}{r}\bigg(\frac{\lambda N^2}{\omega_T\omega}\bigg)^{1/4}\,,
    \eeq
    where $\theta=3\pi/8, 7\pi/8, 11\pi/8\;\mathrm{or}\; 15\pi/8$, corresponding to the four solutions of rapidly increasing, slowly increasing, rapidly decreasing and slowly decreasing thermal modes, respectively (Figure \ref{fig:k_complex_plane}). Physically, these waves are gravito-thermal modes in which both buoyancy and thermal diffusion play important roles. For the rapidly evanescent modes, $\Im(k_r)=\pm(\lambda N^2/\omega_T\omega)^{1/4} \cos(\pi/8)/r$, while for the slowly evanescent modes, $\Im(k_r)=\pm(\lambda N^2/\omega_T\omega)^{1/4} \sin(\pi/8)/r $, both of which are independent of $\gamma$. 
\end{enumerate}

Since $\omega_\mathrm{crit}$ explicitly depends on the local stellar properties, modes of a given frequency can behave as either gravity or thermal waves in different parts of the star, as we see in Figure \ref{fig:eigenfunction}. Modes can therefore behave as ``mixed modes", with gravity mode character in the core of the star where thermal diffusion is unimportant, and thermal mode character near the surface of the star where thermal diffusion is very important. Such modes have rarely been examined in asteroseismology because their high damping rates mean that they will not be visible as stellar pulsation modes. However, these damping rates also mean they could be very important for energy dissipation via tidal excitation.

\section{Strange Modes}
\label{app:strange_modes}

\begin{figure}
    \centering
    \includegraphics[width=\columnwidth]{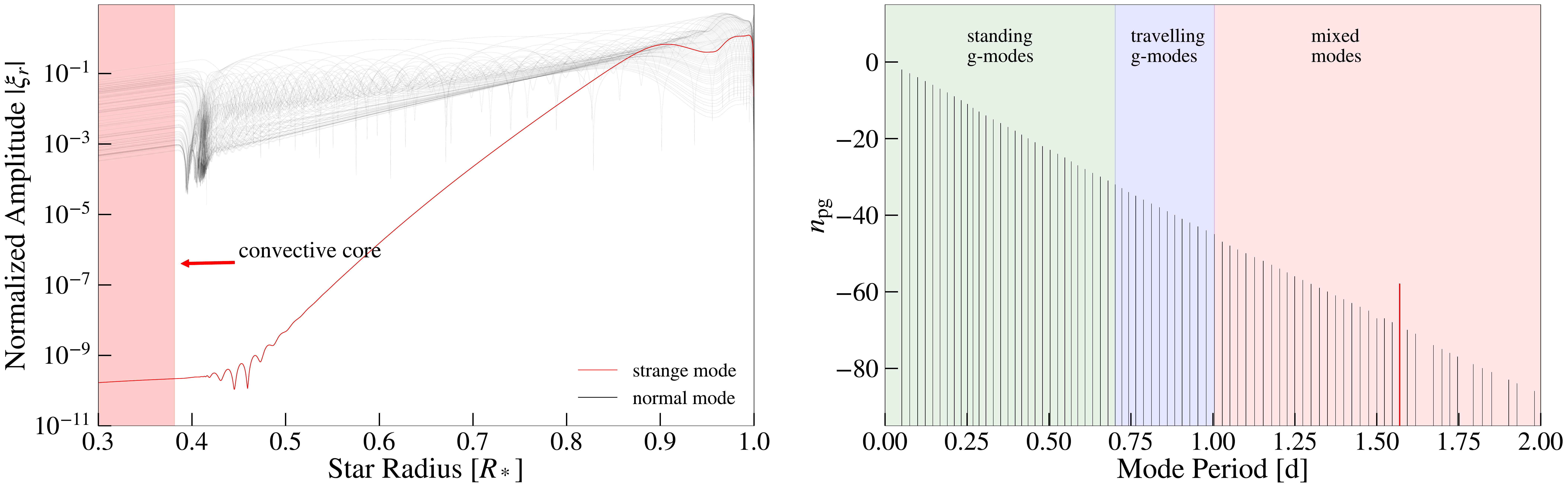}
    \caption{{\bf Left}: All mode eigenfunctions for mode periods between $0.1$ to $2$ days solved by GYRE for a $10\,M_\odot$ Wolf--Rayet star model at solar metallicity during helium burning. In addition to the normal modes (black lines, including standing g-modes, travelling g-modes and mixed modes, discussed in \S \ref{sec:results-modes}), a strange mode solution (red line) exists. This mode has much higher damping rate because it is localized near the stellar surface, as indicated by its rapidly decreasing amplitude towards the core. {\bf Right}: The periods and winding numbers ($n_\mathrm{pg}$) for mode solutions, showing an outlying strange mode solution.}
    \label{fig:strange_mode}
\end{figure}

When solving for high-order, low-frequency mixed modes, GYRE occasionally returns solutions which we identified as ``strange modes''. An example is given in Figure \ref{fig:strange_mode}. The strange modes are usually distinct in the following aspects: 1) the modes have higher damping rates, often one order of magnitude larger than the normal modes. This causes stronger spatial evanescence as the waves propagate inwards, as seen from Equation \ref{eq:krtherm}. 2) The modes have unusual winding numbers $n_\mathrm{pg}$ (defined in \cite{Takata2006}, and treated as mode radial orders in GYRE), departing from the normal $n_\mathrm{pg}$-period relation of normal modes (Figure \ref{fig:strange_mode}, right panel). 3) The modes do not obey the uniform period spacing shared by normal g-modes. 4) The strange mode eigenfunctions seem to be artificially truncated in the radiative envelope, once they reach a minimum amplitude. We confirm that there are no special physical conditions inside the star where they are truncated. Resolution tests also show that the strange mode solutions do not converge even at very high spatial/frequency resolution.

We guess that the strange modes are effectively gravito-thermal mixed modes that are trapped in the near surface region where the waves behave as thermal waves ($|\omega_{\rm crit}| \gg |\sigma|$). Because they are trapped in the surface layers, their damping rates are much larger than normal modes, similar to the acoustic strange modes found at high frequencies \citep{glatzel:02} . Because their eigenfunctions evanesce so rapidly towards the core, their amplitudes apparently drop below the numerical precision of GYRE near the core, causing the artificial behavior of the eigenfunction at small radii seen in Figure \ref{fig:strange_mode}. This also causes the value of $n_{\rm pg}$ computed by GYRE to be incorrect, and explains why their exact frequencies/eigenfunctions do not converge at high spatial resolution. 

In our calculations, the strange modes only exist in the low-frequency range of mode spectra. Hence, if actually physically present, these modes will only be relevant at the late stage of spin--orbit evolution, when the star has already been significantly spun up. Hence, we believe our main results to be robust against the uncertainties surrounding strange modes, but these modes should be studied in more detail in future work.

\section{Nonlinear Damping of Modes}
\label{app:nonlinear}

Our tidal calculations are based entirely on linear theory, yet under certain circumstances nonlinear effects could be important. The dominant nonlinear term in the fluid momentum equation is $\xi\cdot\nabla\xi\sim \xi(d\xi_r/dr)$, hence $d\xi_r/dr$ serves as an approximate measure of linearity, which only holds when $d\xi_r/dr\ll 1$. For our most massive models, the modes become nonlinear very close to resonance, so nonlinear effects will be important during resonance crossings. While developing a complete nonlinear theory is beyond the scope of this work, here we propose an ad-hoc estimate of the nonlinear damping rates of modes.

We have pointed out that the nonlinearity can be estimated by $d\xi_r/dr$. Specifically, we define
\beq
\label{eq:nl-param}
\phi_\alpha \equiv (d\xi_{\alpha,r}/dr)_\mathrm{max} = A_\alpha(d\bar{\xi}_{\alpha,r}/dr)_\mathrm{max}
\eeq
as the parameter to characterize nonlinearity of mode $\alpha$, where $\bar{\xi}_{\alpha,r}$ is the normalized eigenfunction solution of $\alpha$ and $A_\alpha$ is the mode amplitude due to linear driving, given by \cite{fullerheartbeat:17}:
\beq
\label{eq:nl-amplitude}
A_\alpha = \frac12\frac{W_{lm}Q_\alpha\omega_\mathrm{f}}{\sqrt{(\omega_\alpha-\omega_\mathrm{f})^2+(\gamma_\alpha+\gamma_{\alpha,\mathrm{NL}})^2}}\bigg(\frac{M_\mathrm{p}}{M_*}\bigg)\bigg(\frac{R_*}{a}\bigg)^{l+1}\,.
\eeq
Note that we have replaced the damping rate by $\gamma_\alpha+\gamma_{\alpha,\mathrm{NL}}$, where we denote $\gamma_\alpha$ as the usual radiative damping rate we adopted in linear theory, and $\gamma_{\alpha,\mathrm{NL}}$ as the damping rate caused by nonlinear effects. From our convention, $\gamma_{\alpha,\mathrm{NL}}$ is a function of $\phi_\alpha$, i.e. $\gamma_{\alpha,\mathrm{NL}}=\gamma_{\alpha,\mathrm{NL}}(\phi_\alpha)$. While the detailed functional form of $\gamma_{\alpha,\mathrm{NL}}$ requires a thorough examination of the nonlinear damping mechanisms, physically we expect
\beq
\label{eq:nl-requirement}
\gamma_{\alpha,\mathrm{NL}}(0)=0\,,\;\gamma_{\alpha,\mathrm{NL}}(\phi_\alpha\gtrsim 1)\simeq\gamma_\mathrm{\alpha,NL,max}\,,
\eeq
i.e., no nonlinear damping when the mode amplitude is zero, and maximum damping when the $\phi_\alpha$ parameter reaches $1$. The maximum damping rate $\gamma_\mathrm{\alpha,NL,max}$ can be estimated by the inverse group travel time $\tau_{\alpha,2}$ defined in \cite{ma2021}:
\beq
\gamma_\mathrm{\alpha,NL,max}\simeq-\frac{1}{\tau_{\alpha,2}}=-\bigg(\frac{\sqrt{6}}{\omega_\alpha^2}\int_\mathrm{rad}\frac{Ndr}{r}\bigg)^{-1}\,,
\eeq
where the integral is carried out in the radiative zone of the star, and $N$ is the Brunt-V\"ais\"al\"a frequency. Several authors suggest that the nonlinear damping rate should scale as $\gamma_{\alpha,\mathrm{NL}}\propto\sqrt{E_\alpha}\propto A_\alpha$ (see, e.g., \citealt{Kumar1996,yu2020}).  This suggests  $\gamma_{\alpha,\mathrm{NL}}(\phi_\alpha)$ is linearly proportional to the mode amplitude, such that
\beq
\label{eq:nl-damping-rate}
\gamma_{\alpha,\mathrm{NL}}(\phi_\alpha)\simeq\mathrm{min}(1,\phi_\alpha)\gamma_\mathrm{\alpha,NL,max}\,.
\eeq
Note that this ad-hoc expression \ref{eq:nl-damping-rate} should most likely to hold when $\phi_\alpha\ll1$ and $\phi_\alpha\gtrsim 1$, since we only know the properties of this function under these two limits. This further suggests we can assume $(\gamma_\alpha+\gamma_{\alpha,\mathrm{NL}})^2\simeq\gamma_\alpha^2+\gamma_{\alpha,\mathrm{NL}}^2$, since one of the two terms will always dominate the expression under these two limits. With this convention, combining Equations \ref{eq:nl-param}, \ref{eq:nl-amplitude} and \ref{eq:nl-damping-rate}, we have a quadratic equation for $A_\alpha^2$ (when $\phi_\alpha<1$)
\beq
(A_\alpha^2)^2+2B_\alpha A_\alpha^2-C_\alpha=0\,,
\eeq
where $B_\alpha=((\omega_\alpha-\omega_\mathrm{f})^2+\gamma_\alpha^2)/(2\bar{\gamma}_\alpha^2)$, $C_\alpha=W_{lm}^2Q_\alpha^2(\omega_\mathrm{f}/2\bar{\gamma}_\alpha)^2(M_\mathrm{p}/M_*)^2(R_*/a)^{2(l+1)}$ and $\bar{\gamma}_\alpha\equiv (d\bar{\xi}_{\alpha,r}/dr)_\mathrm{max}\gamma_\mathrm{\alpha,NL,max}$. The positive solution of $A_\alpha^2$ gives
\beq
A_\alpha = (\sqrt{B_\alpha^2+C_\alpha}-B_\alpha)^{1/2}\,.
\eeq
Hence the nonlinear damping rate is given by
\beq
\gamma_{\alpha,\mathrm{NL}}=-\mathrm{min} \big[ 1, \big( \sqrt{B_\alpha^2+C_\alpha}-B_\alpha \big)^{1/2}(d\bar{\xi}_{\alpha,r}/dr)_\mathrm{max} \big] \tau_{\alpha,2}^{-1} \,.
\eeq

\bibliography{tidal}{}
\bibliographystyle{aasjournal}

\end{CJK*}
\end{document}